\documentclass[prb,twocolumn,showpacs,preprintnumbers,amsmath,amssymb]{revtex4-1}

\usepackage{graphicx}
\usepackage{dcolumn}
\usepackage{bm}

\begin{document}
\def\3he{$^3$He}
\def\4he{$^4$He}


\title{Mass flow through solid \4he induced by the fountain effect}

\author{M.W. Ray}
\author{R.B. Hallock}%
\affiliation{%
Laboratory for Low Temperature Physics, Department of Physics,\\
University of Massachusetts, Amherst, MA 01003
}%

\date{\today}

\begin{abstract}
Using an apparatus that allows superfluid liquid \4he to be in contact with hcp solid \4he at pressures greater than the bulk melting pressure of the solid, we have performed experiments that show evidence for $^4$He mass flux through the solid and the likely presence of superfluid inside the solid.  We present results that show that a thermomechanical equilibrium in quantitative agreement with the fountain effect exists between two liquid reservoirs connected to each other through two superfluid-filled Vycor rods in series with a chamber filled with solid \4he.  We use the thermomechanical effect to induce flow through the solid and measure the flow rate.  On cooling, mass flux appears near $T = 600$ mK and rises smoothly as the temperature is lowered.  Near $T = 75$ mK a sharp drop in the flux is present.  The flux increases as the temperature is reduced below 75 mK. We comment on possible causes of this flux minimum.
\end{abstract}

\pacs{67.80.bd, 67.80.B-}
\maketitle
\section{Introduction}
In 2004, Kim and Chan \cite{Kim2004a} observed a decrease in the resonant period of a torsional oscillator filled with solid \4he, which they interpreted as likely evidence for a superfluid phase of solid helium.
This observation stimulated renewed interest in the low temperature properties of solid \4he.  Aside from the torsional oscillator experiments, which have been replicated in several different laboratories (for recent reviews see references~[\onlinecite{Prokofev2007,Balibar2008,Galli2008}]), unexpected behavior has also been observed in the shear modulus of solid helium where the solid helium is seen to become stiffer at the same temperature at which the torsional oscillator experiments show a decrease in the resonant period\cite{Day2007}.  Further, the shear modulus increase shows the same dependence on \3he concentration and temperature as the period shift of the torsional oscillators, which suggests that there is a connection between the results of the torsional oscillator and shear modulus experiments.  This connection has been emphasized recently\cite{Pratt2011}. A peak in the heat capacity\cite{Lin2007} of solid helium was also observed, which may indicate that a phase transition takes place in the solid.

It is unlikely that these observations support a ``supersolid" \cite{Mullin1971} phase of the type originally proposed by the early theoretical studies\cite{Andreev1969,Chester1970,Leggett1970}, which were based on the presence of vacancies.  In fact, the possibility that delocalized vacancies form a superfluid in solid helium at very low temperature has been recently shown by theorists to be impossible \cite{Ceperley2004,Clark2006,Prokofev2005,Boninsegni2006a}.  Torsional oscillator measurements made on quench-cooled, highly disordered crystals show a large period shift, while in annealed crystals the effect nearly disappears \cite{Rittner2006}.  Such behavior points to a disorder-driven phenomenon, meaning that the period shift may not be an intrinsic property of the solid.  In fact, some defects, such as grain boundaries \cite{Pollet2007} and dislocations \cite{Boninsegni2007} have been shown in simulations to support a supefluid phase, as has glassy, or amorphous, solid helium \cite{Boninsegni2006,Andreev2007}.  The observed heat capacity peak \cite{Lin2007} may also be explained by a glass transition \cite{Su2010}.  This glass transition theory, which does not invoke the presence of a superfluid phase, is supported by experimental data showing a $T^2$ contribution to the pressure of solid helium as a function of temperature \cite{Grigorev2007a}, as well as the observations of long time constants at lower temperatures \cite{Hunt2009}, both of which can be explained by glassy regions in the solid \cite{Su2010,Su2010a}.

Adding to the complexity of the situation is a torsional oscillator experiment\cite{Reppy2010} that shows disorder may affect the high temperature properties of the solid but not the low temperature properties.  This is consistent with what is observed for the shear modulus, which shows stiffening at low temperatures, but where stressing and annealing the solid affect the high temperature properties of the solid \cite{Day2009}.  These measurements lead to the interpretation that there may be no evidence for supersolid behavior in torsional oscillator measurements.  Rather, it has been suggested that there is elasticity at higher temperatures and as the temperature is lowered elasticity is reduced, which causes the period of solid-filled torsional oscillators to decrease.  Softening of solid \4he has also been seen recently by acoustical techniques and interpreted as consistent with the unbinding of dislocations from \3he impurities\cite{Rojas2010a}.

On the other hand, some experiments seem to support the supersolid interpretation.  Recent, simultaneous torsional oscillator and shear modulus measurements appear to show a suppression of the resonant period drop of the torsional oscillator under D.C. rotation\cite{Choi2010}, with no rotation effect on the shear modulus measurements on the solid in a different location in the same apparatus.  These simultaneous measurements made under rotation have been interpreted\cite{Choi2010} as evidence for the existence of a superfluid phase in the solid.  Other experiments that may provide evidence for a supersolid phase are (1) the torsional oscillator experiments that used Vycor\cite{Kim2004a} or porous gold\cite{Kim2005}, which seem hard to explain by the presence of dislocations, and (2) the reduced signal seen in a blocked-annulus experiment\cite{Kim2004b}.  However, a different blocked annulus experiment by Rittner and Reppy, unpublished but noted in ref ~[\onlinecite{Reich2010}], provided results that complicate the picture.

If solid helium does indeed support superfluidity, then it should be possible to induce mass flow in solid helium and one would expect to find a flow-limiting critical velocity.  Such a flow has been seen by injecting mass into one side of the solid, and observing a pressure relaxation in a reservoir on the other side\cite{Ray2008a,Ray2009b} for $T \lesssim 550$ mK.  The mass flux was seen to be independent of the driving pressure difference, which is consistent with a superflow at critical velocity.  Frost heave\cite{Hiroi1989}, plastic flow, or liquid channels\cite{Sasaki2007,Sasaki2008,Ray2008e} that might provide alternate explanations for the data that show mass flux through the solid seem inconsistent with the behavior of the flux as a function of pressure and temperature.  Those flow experiments were different in various respects from other flow experiments \cite{Bonfait1989,Greywall1977,Day2005,Day2006,Sasaki2006,Rittner2009} in that they were simultaneously (1) performed at pressures greater than the bulk melting pressure, and (2) done by injecting mass into the solid, and not by trying to push bulk solid through a constrained geometry.  Experiments which have attempted to mechanically push solid $^4$He  through small-dimensioned passages have never shown any mass flow \cite{Greywall1977,Day2005,Day2006,Rittner2009}; squeezing the lattice directly does not result in mass flow.

Here we use an apparatus similar to that used in our previous flow experiments \cite{Ray2008a,Ray2009b}, which allows us to have superfluid liquid helium in contact with the solid at a pressure greater than the bulk melting pressure of solid helium. We present two main results, which were briefly reported on previously \cite{Ray2010b,Ray2010c}, that support the possibility that superfluidity exists in solid helium.  In section \ref{sec:concept} we describe the concept of the experiments and describe in detail the apparatus used.  In section \ref{sec:fountain} we present results that show that when a temperature difference is imposed between two reservoirs connected by Vycor rods in series with a chamber of solid helium, a pressure difference is induced between the two reservoirs that is described well by the thermo-mechanical (fountain) effect \cite{wilks}. In section \ref{sec:flux}, we show the results of experiments in which we use the fountain effect to induce flow through the solid.  We describe the behavior of the flow at high temperature, and at low temperature where we observed an unexpected drop in the flux, followed by an increase in flux at the lowest temperatures we could reach with our apparatus ($\approx$ 60 mK).  In section \ref{sec:disc} we discuss our results and offer comments. Finally, in section \ref{sec:sum} we summarize our results and present some further comments and conclusions.

\section{Concept and design}
\label{sec:concept}
\begin{figure}
\resizebox{3 in}{!}{
\includegraphics{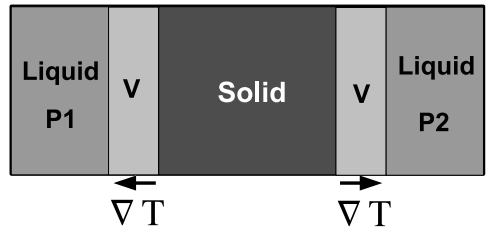}}
\caption{\label{fig:concept} Conceptual design of the flow experiments.  Liquid-filled Vycor separates the solid from the two liquid reservoirs, and a temperature gradient across the Vycor allows the liquid reservoirs to remain in the bulk liquid region of the phase diagram, while the solid can be cooled to lower temperatures.  The elevation of the melting pressure for the helium in the Vycor allows for an interface between the superfluid in the Vycor, and the bulk solid.}
\end{figure}
\begin{figure}
\resizebox{3 in}{!}{
\includegraphics{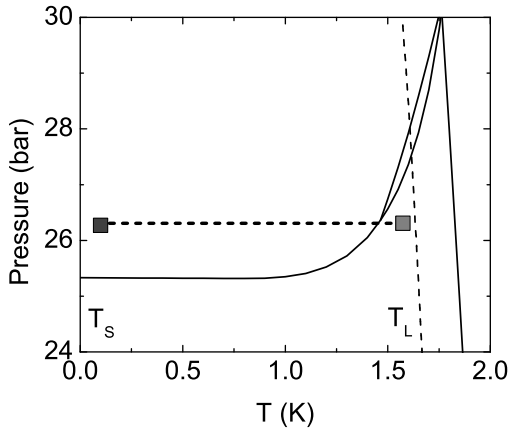}}
\caption{\label{fig:vphase-exp} Phase coordinates of the liquid reservoirs at temperature $T_L$ and the solid at a typical temperature $T_S$ and pressure where the experiments are preformed.  The dashed horizontal line shows the temperature gradient along the Vycor. The dashed nearly vertical line represents the approximate location of the Lambda transition in helium-filled Vycor.}
\end{figure}

The general design for the apparatus is similar to that used previously to search for D.C. flow through solid helium by mass injection \cite{Ray2008a,Ray2009b}.  It exploits the properties of helium in confined geometries to allow for an interface between superfluid liquid and solid \4he at pressures greater than the bulk melting pressure of solid helium.   The concept of the experiment is shown in figure \ref{fig:concept}. It is known \cite{Beamish1983,Lie-zhao1986,Adams1987} that the liquid-solid transition for helium in porous Vycor glass is elevated so that when the pressure of the helium in the Vycor is less than $\approx 35$ bar, it remains a liquid.  A chamber filled with solid \4he is sandwiched between two liquid chambers with helium-filled Vycor separating the solid from the liquid.  By imposing a temperature gradient across the Vycor, the outer chambers can be kept in the bulk liquid (superfluid) region of the phase diagram with the center chamber in the bulk solid region (figure \ref{fig:vphase-exp}). Thus an interface can exist between the superfluid liquid $^4$He in the Vycor  and the solid in the center chamber at pressures greater than the bulk melting pressure of solid helium ($\approx 25$ bar).  In the absence of a porous material such as Vycor, the liquid-solid interface can only occur at the bulk freezing-melting pressure of the solid.  Our design allows a chemical potential difference to be imposed across the solid with the Vycor acting as electrodes.

\begin{figure}
\resizebox{3 in}{!}{
\includegraphics{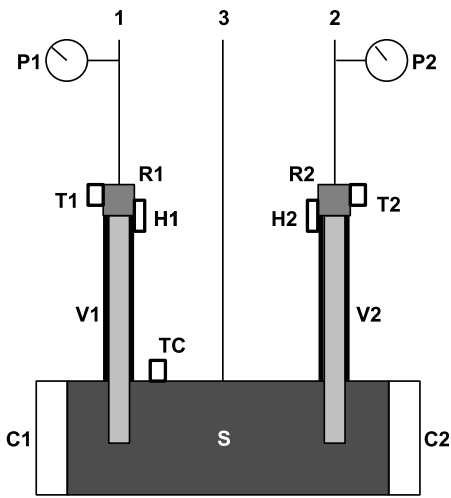}}
\caption{\label{fig:cell} Schematic diagram of the cell used for
flow experiments.  Three fill lines lead to the cell, capillaries 1 and 2 go to
liquid reservoirs R1 and R2 above the Vycor Rods V1 and V2.  The
third fill line leads directly to the solid chamber, S.  Two capacitance
pressure gauges, C1 and C2, are located on either side of the cell for
\textit{in situ} pressure measurements.  Pressures in the Vycor
lines (1 and 2) are read by pressure transducers outside of the
cryostat.  Each reservoir has a heater, H1 and H2, which prevents
the liquid in it from freezing, and the reservoir temperatures are
read by calibrated carbon resistance thermometers T1 and T2.  The cell
temperature is recorded by a third calibrated carbon resistance thermometer,
TC. The cell thermometer reading, denoted TC, measures the
temperature of the sample, T.}
\end{figure}
A schematic diagram of the cell used for these experiments is shown in figure \ref{fig:cell}.  The solid helium is grown in region S, which has a cylindrical geometry, dia = 0.635 cm, (with side regions for access for future experiments), with a volume $V_{cell} = 1.84$ cm$^3$.  A capacitance strain gauge of the Straty and Adams type \cite{Straty1969} (C1 and C2) is on each end of region S for {\it in situ} measurement of the pressure of the solid.  Two Vycor rods, V1 and V2, enter region S through the top of the cell.  The Vycor rods are 0.140 cm in diameter, 7.620 cm in length, and the cylindrical surface of the Vycor external to region S is sealed with a thin coating of Stycast 2850 FT epoxy.  The Vycor penetrates into the cylindrical cell to about the mid-line axis of the cylinder.  At the top of each Vycor rod is a liquid reservoir, R1 and R2; each reservoir is temperature-controlled with a heater, H1 and H2.  The bottom of each Vycor rod has an epoxy patch across the end of the rod to ensure that an anomaly\cite{Ray2009b,Wilson1968} typically found along the axis of a Vycor rod is not an issue.  Thus the cell is akin to a U-tube with solid at the bottom in chamber S, and superfluid on the sides in the Vycor and reservoirs.

The reservoirs are fed by stainless steel capillaries 1 and 2, which are heat sunk only at 4 K.  To aid in the temperature control, each reservoir is thermally connected to the still of the dilution refrigerator, which is typically operated at $T \approx 800$ mK.  The pressures in these lines, denoted as $P1$ and $P2$, are measured by pressure transducers located at room temperature.  Another capillary, line 3, is used to fill the cell initially. This line is heat sunk at several places on the refrigerator, bypasses the Vycor rods and leads directly to region S.  The cell is mounted on a copper plate which is attached to the mixing chamber of a dilution refrigerator by six copper rods of diameter 0.635 cm.  The pressure gauges P1 and P2 are Paroscientific Digiquartz pressure transducers and the capacitive gauges C1 and C2 are monitored by an Andeen-Hagerling 2500-A capacitance bridge (C1) and by a General Radio 1615A bridge (C2), with a Stanford Research Systems SR830 lock-in amplifier.  The reservoirs R1 and R2 shown in figure \ref{fig:cell} represent the outer liquid chambers shown in figure \ref{fig:concept}.  The heaters create the temperature gradient along the Vycor rods.  Since capillaries 1 and 2 (which lead to the reservoirs) are heat sunk only at 4 K, no solid can form in them, and this provides a direct connection from the helium supply at room temperature to the reservoirs. With solid helium in region S, there is also solid helium in line 3, which prevents mass flow into or out of the cell through this capillary.

Mass flow in a superfluid system is governed by the chemical potential difference, $\Delta \mu$.  If solid helium supports a superfluid flow, then flow should be induced by the imposition of $\Delta \mu$ between the Vycor rods.  In the present context, for a superfluid $\Delta \mu = v \Delta P - s \Delta T$ where $v = V/N$ is the volume per particle and $s = S/N$ is entropy per particle.  This implies that a flow can be induced by either a pressure difference\cite{Ray2008a,Ray2009b}, $\Delta P$,  or a temperature difference\cite{Ray2010c}, $\Delta T$.  Previously we induced flow with the former.  Here we focus on the latter. For such experiments, once a solid is present in the cell lines 1 and 2 are valved closed and line 3 is blocked by solid helium.  By changing the thermal energy supplied to H1 and/or H2, we impose a temperature difference between the two reservoirs and monitor the flow between the two reservoirs by measuring the rate at which the pressure changes in the reservoirs, $dPi / dt$ where $i = 1, 2$ for the two pressure gages.

The \4he we use is ultra high purity \4he with an assumed \3he impurity concentration of $\approx$ 300 ppb; it was passed though a liquid nitrogen and a liquid helium trap.  All of our samples are grown in region S, typically from the superfluid at constant temperature (for a more detailed discussion of this growth method, see ref.~\onlinecite{Ray2010a}), but a few samples have been grown by the blocked capillary method\cite{Ray2009b}.  We note here that we have previously discussed\cite{Ray2010c} the fact that growth above the melting curve from the superfluid has interesting temperature dependence, with such growth not possible for fresh samples below $\approx$ 300 mK.  But, samples can be grown by this method above 300 mK, then be cooled and studied at lower temperature.

Once the solid is grown in the chamber, we can search for mass flow through the solid by changing the amount of thermal power deposited in the heaters H1 and H2.  This procedure differs from the injection method \cite{Ray2008a,Ray2009b}, because after each measurement, the heater powers can be returned to their original settings, thus returning the system to the same initial conditions of temperature and pressure for the next measurement. In the injection method if the injection resulted in mass flow, the injection increased the pressure of the solid, and the subsequent measurement was then done at a higher pressure.

Our solid samples are identified by pairs of letters, assigned to the samples or to measurements of the samples in chronological order.  So, for example, a sample might be created and labeled as GO and its pressure might be increased and the sample renamed as GP. Subsequent flow attempts at different temperatures on the same sample might be labeled GQ, GR, etc. By this means each measurement or manipulation of a sample is assigned a unique code.  Tabulation of all solid samples discussed here is presented in the Appendix.

\section{equilibrium fountain pressure}
\label{sec:fountain}
If two containers of superfluid $^4$He are connected by a superleak, a path through which only superfluid can flow, then a temperature difference imposed between the two containers will give rise to a pressure difference.  This pressure difference, $\Delta P_f$, known as the fountain pressure, is given by the fountain equation \cite{wilks}
\begin{equation}
\Delta P_f = \int_{T_a}^{T_b} \rho S dT.
\label{eq:ftn}
\end{equation}
Here $\Delta T = T_a - T_b$ is the temperature difference between the two containers, and $\rho$ and $S$ are the density and entropy of the liquid, respectively.  When needed, values for $\rho(P,T)$ and $S(P,T)$ were taken from Maynard\cite{Maynard1976} and Donnelly\cite{Donnelly1967} and extrapolated to above 25 bar where necessary.

\begin{figure}
\resizebox{3in}{!}{
\includegraphics{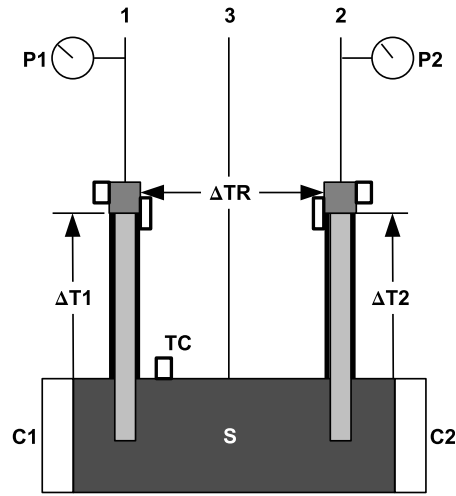}}
\caption{\label{fig:cell-DT} Temperature gradients in the cell ($\Delta T1$, $\Delta T2$
and $\Delta TR$, which lead to fountain pressures.  }
\end{figure}
 With reference to figure \ref{fig:cell-DT} there are three fountain pressures to consider. One is due to the temperature difference between R1 and the cell, $\Delta T1 = T1 - TC$,  another is due to the temperature difference between R2 and the cell, $\Delta T2 = T2 - TC$, and finally there is $\Delta TR$ = $T1 - T2$, due to the temperature difference between R1 and R2.  With superfluid present, $\Delta T1$ and $\Delta T2$ lead to separate pressure differences between the cell and the respective liquid reservoir, $\Delta P1$ or $\Delta P2$, given by equation \ref{eq:ftn}, which are determined solely by the temperature of the respective reservoirs and $TC$.  A non-zero value for $T1 - T2$ results in a pressure difference between the two reservoirs, $\Delta PR = P1 - P2$, which can also be found using equation \ref{eq:ftn} with $T_a = T1$ and $T_b = T2$.  If there is a supersolid that allows superflow in the cell, one may also think of $\Delta PR$ in a different way. In that case, one may think of the superfluid in R1 and R2 as connected by a superleak, V1+S+V2.  With this perspective, we also have $\Delta PR = P1 - P2$.

\subsection{equilibrium fountain pressure in liquid helium}
\begin{figure}
\resizebox{3.5in}{!}{
\includegraphics{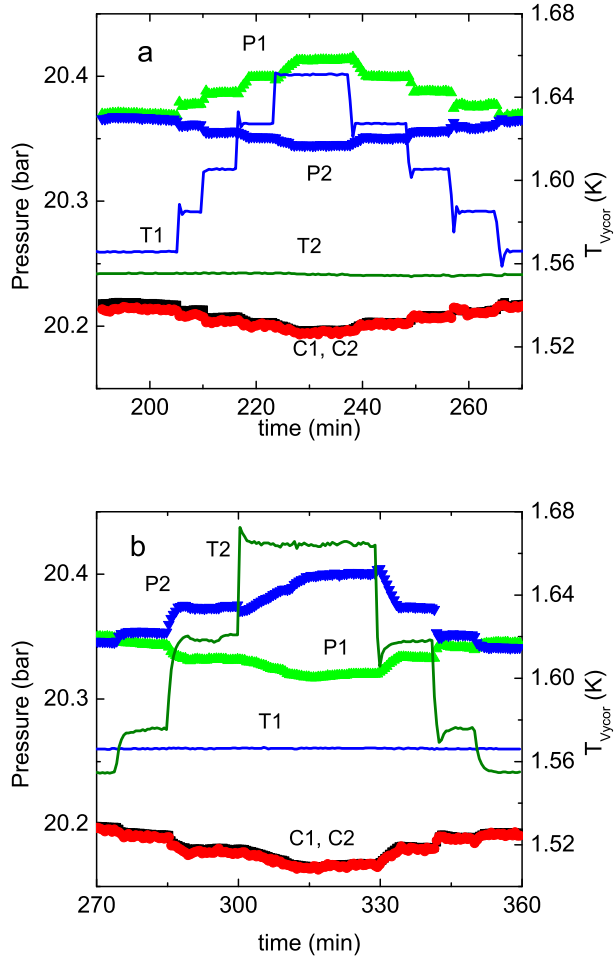}}
\caption{\label{fig:lhe-ftn} (color online) Measurement of the fountain effect in liquid $^4$He.  Each increase in a reservoir temperature is accompanied by an increase in the pressure of the accompanying reservoir and a decrease in the pressure of the other reservoir.  Decreasing the temperature reverses the process.  (a) Changes in T1, (b) changes in T2.  For these data $TC$ = 65 mK.  Transient effects in $T1$ and $T2$ associated with changes in temperature are due to the operation of the temperature regulators.}
\end{figure}
Figure \ref{fig:lhe-ftn} shows the results of measurements that document the presence of the fountain effect when {\it liquid} $^4$He fills chamber S at 65 mK near 20 bar.  In figure \ref{fig:lhe-ftn}a, T2 is held constant at 1.555 K while T1 is changed.  First, T1 is raised in steps by increasing the thermal power deposited by H1.  Each time $T1$ is raised, a rise in $P1$ is recorded with a corresponding drop in the cell pressure ($C1$ and $C2$) and $P2$.  Next, when T1 is lowered by decreasing the thermal power deposited by H1, $P1$ is seen to drop, while $P2$, $C1$ and $C2$ all rise.  When $T1$ is returned to its original value, all pressure values return to their original values; there is no hysteresis.   Figure \ref{fig:lhe-ftn}b shows the same type of measurement only this time $T1$ is held constant at 1.566 K and $T2$ is changed.  Note that in figures \ref{fig:lhe-ftn}a,b for relatively large values of $T1$ or $T2$ (e.g. $T2$ in figure \ref{fig:lhe-ftn}b) the flow rate slows as the temperature of the reservoir is raised.  This is likely due to a decreased superfluid fraction in the upper (warmer) region of the relevant Vycor rod. As will be described, during our flow measurements with solid in the sample cell changes in T1 or T2 were typically limited to $\approx$ 24 mK, as will be evident in figures presented later in this report.

\begin{figure}
\resizebox{3.5in}{!}{
\includegraphics{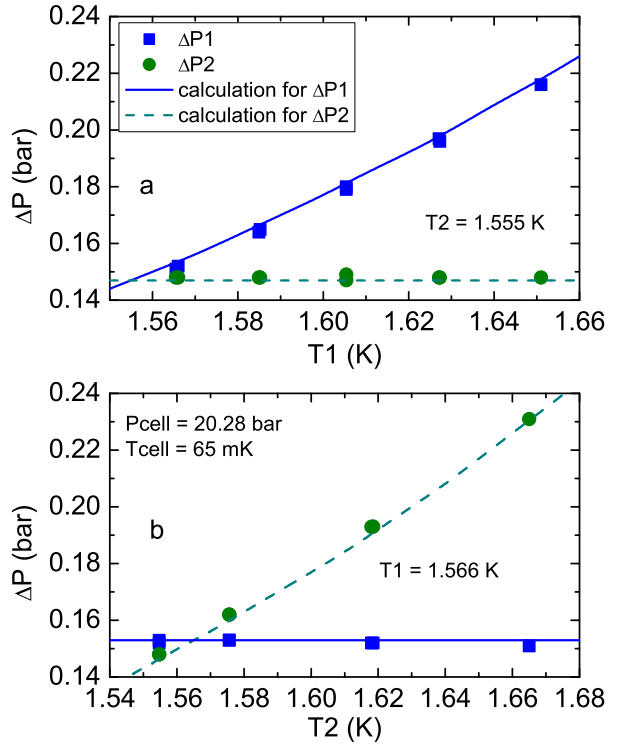}}
\caption{\label{fig:lheftn_T-TC} (color online) Measured pressure difference between the reservoirs and the cell for (a) changes in T1 with T2 held constant, R1, $\Delta P1$ $>$ 0, and (b)changes in T2 with T1 held constant,  R2, $\Delta P2$ $>$ 0, from the data shown in figure \ref{fig:lhe-ftn}, along with the calculated fountain pressure from equation \ref{eq:ftn} for the same conditions.}

\end{figure}
\begin{figure}
\resizebox{3.5in}{!}{
\includegraphics{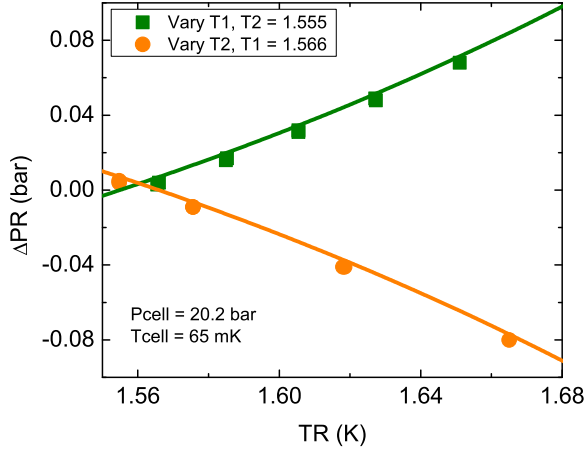}}
\caption{\label{fig:lheftn_T1-T2} (color online) Measured pressure difference between the two reservoirs, $\Delta PR = P1 - P2$, from the data shown in figure \ref{fig:lhe-ftn}, along with the calculated fountain pressure for the same conditions.  In each case TR is the temperature of the reservoir that is not held constant.}
\end{figure}

Figure \ref{fig:lheftn_T-TC} shows the measured pressure difference between each reservoir and the cell (where $C1 = C2$), $\Delta P1 = P1 - C1$ and $\Delta P2 = P2 - C2$, from the data in figure \ref{fig:lhe-ftn} along with the expected fountain pressure using equation \ref{eq:ftn} as described above.  There is quantitative agreement between the two.  Notice that in figure \ref{fig:lheftn_T-TC}a  $\Delta P2 = const.$ and in \ref{fig:lheftn_T-TC}b $\Delta P1 = const.$  This constant behavior can be understood since the temperature difference between the respective reservoir and cell is constant, so the fountain pressure should not change.  Thus, for example, as P1 increases, P2 and the cell pressure (measured {\it in situ} by C1 and C2) decrease together.  Finally, figure \ref{fig:lheftn_T1-T2} shows the pressure difference between the two reservoirs, $\Delta PR = P1 - P2$, using the data shown in figure \ref{fig:lhe-ftn} along with the expected fountain pressure found from equation \ref{eq:ftn}.  Again, there is good agreement between the two meaning that the observed pressure differences observed with liquid \4he in the cell are as expected from the fountain effect, equation \ref{eq:ftn}.

\subsection{equilibrium fountain pressure in the presence of solid}
\begin{figure}
\resizebox{3.5in}{!}{
\includegraphics{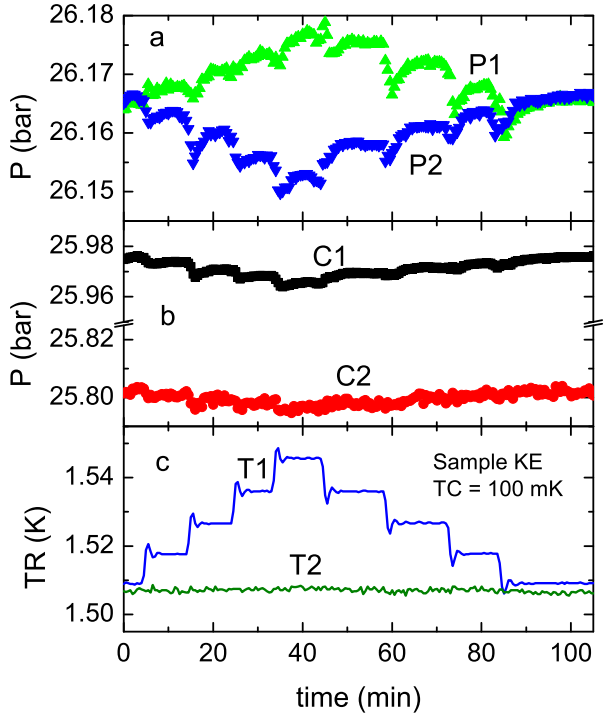}}
\caption{\label{fig:KE} (color online) Solid helium sample KE, $TC = 100$ mK.  (a) Pressures, P1, P2, (b) Pressures C1, C2, (c) reservoir temperatures, T1, T2.  T2 was held constant while T1 was changed in steps to measure the fountain pressures present.  Typically the capacitor monitored with the General Radio bridge, here C2, showed a bit more noise.}
\end{figure}

Figure \ref{fig:KE} shows data from solid sample KE, grown from the superfluid at $TC = 352$ mK and then cooled to 100 mK.  In a similar fashion to the liquid-only case shown in figure \ref{fig:lhe-ftn}, T1 was increased in steps and then subsequently decreased, while T2 was held constant.  Here, each increase in T1 is accompanied by an increase in P1, (as shown for the liquid-only case in figure \ref{fig:lhe-ftn}), and a decrease in P2 and the cell pressures measured by C1 and C2.

\begin{figure}
\resizebox{3.5in}{!}{
\includegraphics{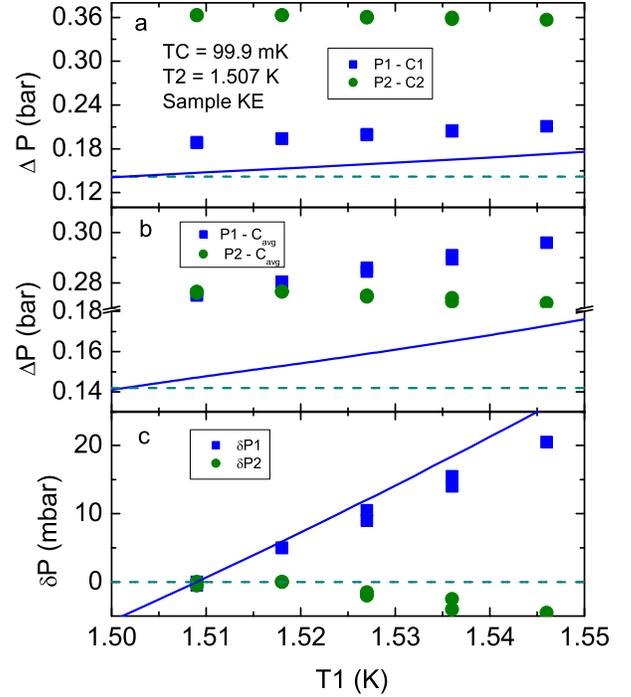}}
\caption{\label{fig:KE-T-TC} (color online) Measurements of the pressure difference between each reservoir and the cell with solid in region S: sample KE, $TC$ = 100 mK (figure \ref{fig:KE}).  In each case the squares and circles refer to reservoirs R1 and R2, respectively, while the solid and dashed lines show predictions based on equation \ref{eq:ftn} for R1 and R2, respectively.  (a) $P1 - C1$ and $P2 - C2$, (b) $P1 - C_{avg}$ and $P2 - C_{avg}$ where $C_{avg} = (C1 + C2) / 2$ is the average between C1 and C2.  Note the axis break.  (c) $\delta P1$ and $\delta P2$ as defined in the text.}
\end{figure}
Although changes in $T1 - T2$ result in changes in $P1$ and $P2$, figures 9a and 9b show that quantitatively the pressure differences induced, $\Delta Pi = Pi - Ci$ are not described by equation 1.  This is not surprising.  As we have noted previously\cite{Ray2009b,Ray2010a}, and as is evident in figure \ref{fig:KE}, there are often cases for which $\Delta C = C1 - C2 \neq 0$ when solid fills the cell. The presence of two separated {\it in situ} pressure gauges shows that the solid helium in our apparatus can sustain a stable pressure gradient; and, presumably the pressure can vary spatially in non-uniform ways throughout the solid.  In fact, even with $C1 = C2$ with solid in the cell there is no reason to expect that the pressure difference $Pi - Ci$ can be predicted using \ref{eq:ftn}. This is because the local pressure measured by each of the capacitors is not the relevant pressure to determine the fountain pressure.  Figure \ref{fig:KE-T-TC} supports this. It shows $Pi - Ci$ (figure \ref{fig:KE-T-TC}a) and $Pi -C_{ave}$ (figure \ref{fig:KE-T-TC}b), where $C_{avg} = (C1 + C2) / 2$ is the average between C1 and C2 from the data shown in figure \ref{fig:KE}, along with the predicted fountain pressure from equation \ref{eq:ftn}.  In this case, the data shows that the measured pressure difference differs from the predictions of equation \ref{eq:ftn}.

We can look instead at the {\it change} in the fountain pressure that results from changing the temperature in the reservoir, $\delta Pi = \Delta Pi^j - \Delta Pi^0$.  Here $\Delta Pi^j$ represents the difference between the pressure in reservoir, $i = 1, 2$, and the average pressure in the cell ($C_{avg} = (C1 + C2) / 2$) after the $j$th temperature step.  $\Delta Pi^0$ is the pressure difference between the reservoir and the average of $C1$ and $C2$, when $T1 = T2$ (i.e. before the temperature step).  Using this approach, $\delta Pi^j$ was determined from the data shown in figure \ref{fig:KE}, and is plotted against $T1$ in figure \ref{fig:KE-T-TC}c, along with the expected values calculated from equation \ref{eq:ftn}.  This tells us that {\it changes} in the cell pressure respond to {\it changes} in the imposed chemical potential in a predictable way. This behavior seems consistent with the presence of the climb of edge dislocations\cite{Soyler2009}.

\begin{figure}
\resizebox{3.5in}{!}{
\includegraphics{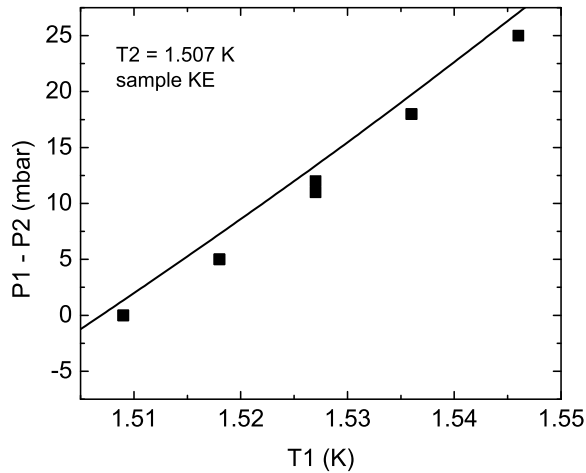}}
\caption{\label{fig:KE-DPR} Measurements of the pressure difference between the two reservoirs, $\Delta PR = P1 - P2$ with solid in region S: sample KE, $TC = $100 mK (figure \ref{fig:KE}).  The expected fountain pressure between the two reservoirs found from equation \ref{eq:ftn} is shown by the solid line.}
\end{figure}
Figure \ref{fig:KE-DPR} shows the equilibrium pressure difference between R1 and R2, $\Delta PR$, determined from the data shown in figure \ref{fig:KE} along with the expected pressure difference computed from equation \ref{eq:ftn}.  Here the data agree well with the fountain effect prediction.  This shows that the presence of lattice pressure gradients in the solid has no effect on the equilibrium pressure difference between the two reservoirs, R1 and R2, caused by the fountain effect. The large pressure gradient in the solid seen in in the case of sample KE (figure \ref{fig:KE}) does not effect $\Delta PR$.  As long as there is mass flow present through the solid helium, any change in the temperature difference between the two reservoirs results in flow through the solid until the chemical potential equilibrates, regardless of any pressure gradients in the solid. Evidence for this is consistently present in the samples we have studied.

\section{measurement of mass flux using the fountain effect}
\label{sec:flux}
\subsection{procedure}
\label{sec:flux:proc}
\begin{figure}
\resizebox{3.5in}{!}{
\includegraphics{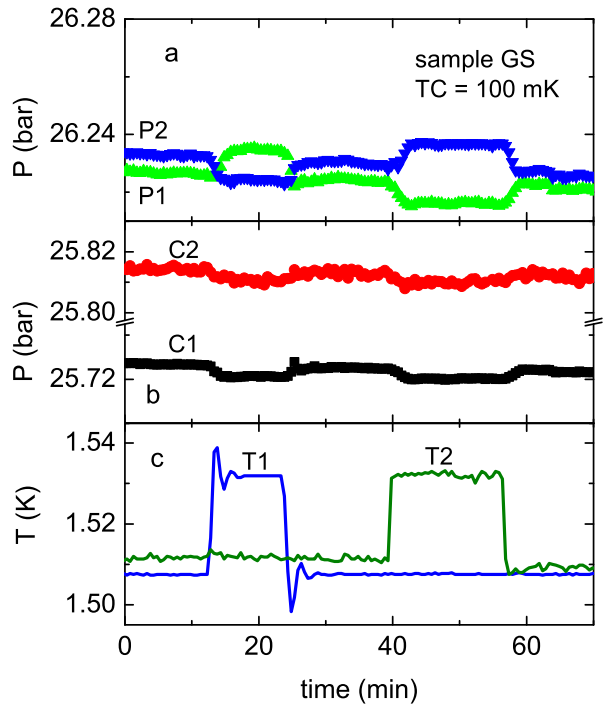}}
\caption{\label{fig:GS} (color online) Data that illustrates the procedure used to measure the flux through solid helium.  (a) Pressures, P1, P2, (b) Pressures C1, C2, (c) reservoir temperatures, T1, T2.  First T1 is raised by $\approx$ 24 mK.  After equilibrium is achieved, T1 is lowered back to its original value, and the procedure is repeated with T2.}
\end{figure}
We next turn to the measurements of the flux induced through the solid when a temperature gradient is imposed between the two reservoirs.  Figure \ref{fig:GS} is an example of the procedure used to measure such flow; sample GS.  This sample was grown from the superfluid at $TC = 317$ mK, cooled to 61 mK (sample GR), then warmed to 100 mK (sample GS, shown here).  Starting with $T1 \approx T2$ ($T1$ is not always precisely equal to $T2$), T1 is raised by $\delta T1 = 24$ mK by increasing the thermal power supplied by H1 while $T2$ is held constant.  $T1$ is held at this new temperature until after the new equilibrium is achieved after which $T1$ is returned to its original value.  Next, after waiting for equilibrium, T2 is changed, with T1 held constant.  For each change in temperature, the mass flux determined by $dP/dt$ is measured to be constant in time with equilibration times of $dt_{eq} \approx 2-3$ min.

\begin{figure}
\resizebox{3.5in}{!}{
\includegraphics{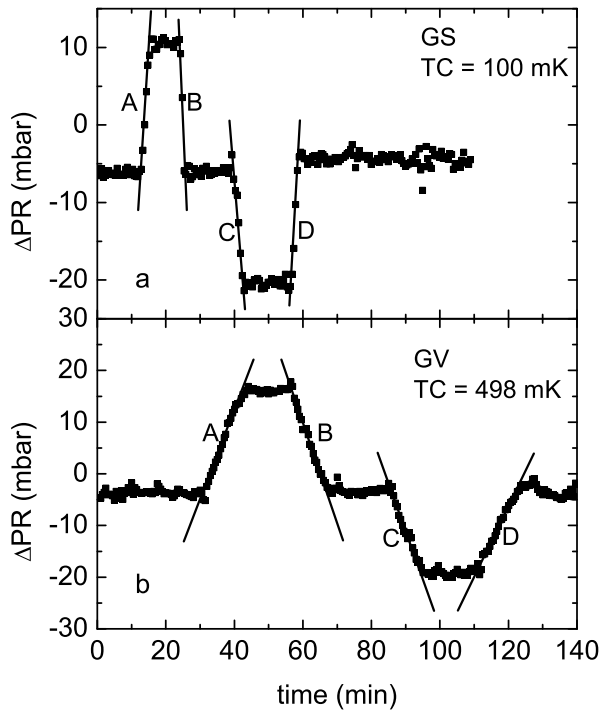}}
\caption{\label{fig:GS-GV_DPR} $\Delta PR = P1 - P2$ for (a) sample GS at $TC = 100$ mK, (from figure \ref{fig:GS}), and (b) sample GV at $TC = 498$ mK.  The non-zero baseline values of $\Delta PR $ result from small baseline differences between $T1$ and $T2$.}
\end{figure}
The constant (in time) nature of the flux following a change in the temperature in one of the reservoirs is more evident in figure \ref{fig:GS-GV_DPR}, which shows the difference between the reservoir pressures $\Delta PR = P1 - P2$.  $\Delta PR$ for sample GS ($TC = 100$mK) is shown in figure \ref{fig:GS-GV_DPR}a.  For comparison we also show $\Delta PR$ for sample GV at $TC = 498$ mK in figure \ref{fig:GS-GV_DPR}b.  The flux for sample GV was much lower than for sample GS, with equilibrium occurring at $t \approx 15$ min.~after changing the reservoir temperature.

A measured quantity proportional to the mass flux through the solid is the slope, $d (\Delta PR) / dt $, that results from changing the reservoir temperature.  We label the slopes A, B, C and D so that, as shown in figure \ref{fig:GS-GV_DPR}, slope A corresponds to the flux observed after increasing T1, and slope B corresponds to the slope that results from reducing T1 back to the original temperature.  Likewise the slope labeled C gives a measure of the flux that results from increasing T2, while slope D results from reducing T2 back to the original temperature.  With these definitions, there will be no ambiguity and all slopes will be reported here as positive numbers.
Given the geometry of our apparatus, a typical value for $d(\Delta P)/dt $ with solid in our cell of $\approx$ 0.05 mbar/sec corresponds to a mass flux of $\approx$ 2.4 $\times 10^{-8}$ g/sec, which is a flux of $\approx$ 3.6 cm$^3$/year at the typical \4he density of these measurements. We report our flux values in the directly determined units, mbar/sec.

Using the procedure detailed above (figures \ref{fig:GS} and \ref{fig:GS-GV_DPR}) the flux through solid helium can be measured at a series of temperatures.  We grow a sample either from the superfluid at constant temperature or by the blocked capillary method.  We then make flux measurements similar to those shown in figures \ref{fig:GS} and \ref{fig:GS-GV_DPR}, then change the temperature of the solid and repeat the measurement at the new temperature.  As mentioned previously, this gives us measurements of the flux at different temperatures with the other parameters of the solid (i.e. base cell pressure) held constant (aside from modest pressure drifts occasionally seen).

\subsection{flow in the range 100 $\leq T \leq $ 700 mK}
\label{sec:flux:highT}
The samples labeled GS (shown in figures \ref{fig:GS} and \ref{fig:GS-GV_DPR}) and GV (shown in figure \ref{fig:GS-GV_DPR}) were part of a larger series of measurements, denoted as series sh03 (see Appendix).  Here, the solid sample was grown from the superfluid at $TC = 317$ mK to a pressure of $C_{avg} = (C1 + C2)/2 = 25.8$ bar.  The sample was then cooled to $TC = 61$ mK, and the flux was measured as detailed above (sample GR).  The temperature was then raised to $TC = 100$ mK (sample GS), then sequentially to $698$ mK (samples GT - GX) with the flux measured after each change in TC of $\approx 100$ mK.  The sample was then cooled back to 60 mK (samples GY - HE), again in $\approx$ 100 mK increments.  Following the measurement at 60 mK the pressure of the solid was increased by a syringe injection \cite{Ray2010a} (i.e. adding atoms through capillaries 1 and 2) to $C_{avg} = 26.0$ bar, and the temperature cycle was repeated (samples HG - HT, series sh04).  More specific details will be provided in section \ref{sec:flux:sample}.  In this section, we will focus on the data for the temperature range $TC \geq 100$ mK.

\begin{figure}
\resizebox{3.5in}{!}{
\includegraphics{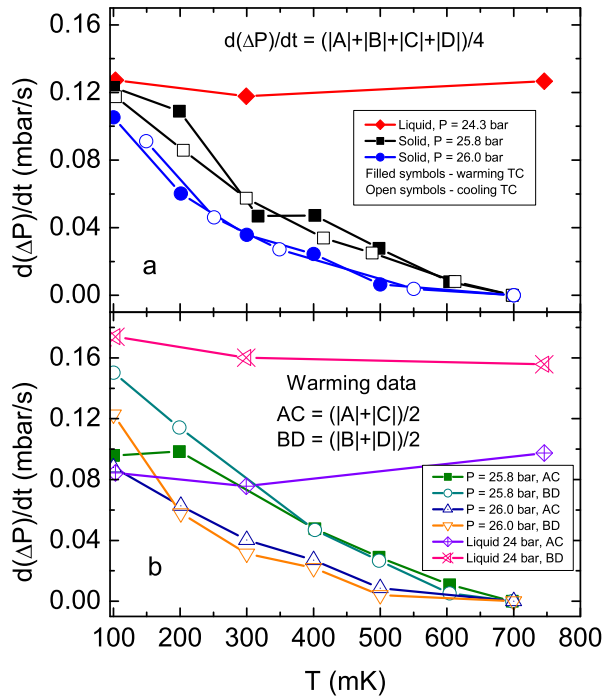}}
\caption{\label{fig:flow-highT} (color online) Measured flux between 100 and 700 mK for series sh03 (25.8 bar), sh04 (26.0 bar), and liquid at 24.3 bar.  (a) average of slopes A, B, C and D showing both warming and cooling TC.  (b) average of A,C and B,D for warming TC only (see text). The liquid-only flux values are also shown.  We take these to indicate that the Vycor may limit the flux, but only in the vicinity of 100 mK. }
\end{figure}

To characterize the flux and make comparisons among samples, we typically average the absolute value of all four slopes and define $d(\Delta P)/dt = ABCD = (|A| + |B| + |C| + |D|)/4$.  Of course, there are other ways to represent the data, which will be discussed later, but here this will be useful and will allow us to bring forward points that we wish to emphasize.
The results of such measurements of the flux found between 100 and 700 mK for the two series described above, series sh03 (samples GR - HE) and series sh04 (samples HG - HT) are shown in figure \ref{fig:flow-highT}a  along with measurements made in the liquid at a pressure of 24.3 bar.  As the temperature is increased from 100 to 700 mK the flux with solid present drops smoothly, with $|d(\Delta P) / dt| = 0$ at $TC = 700$ mK.  Upon cooling from this temperature, the flux is seen to rise, replicating the data taken while warming the solid. Note that series sh04, at a higher pressure, had slightly lower flux.  This is consistent with our previous observations that have shown the flux to decrease with increasing pressure \cite{Ray2009b}.

The liquid data shown in figure \ref{fig:flow-highT}a  represents a measure of the maximum average flux that the Vycor allows at 24.3 bar.  It is seen to be relatively independent of temperature; a similar flux was measured at 22.1 bar. Comparing the flux through the solid to the flux through the liquid, it can be seen that at the higher temperatures, the flux through the Vycor with only liquid in the cell is considerably higher than the flux that is observed when solid is present in the cell.  This suggests that the flux is most likely not limited by the flux through the Vycor in most of this temperature regime.  As the temperature is lowered, the measured flux through the solid increases toward the measured flux through the Vycor with only liquid in the cell.  At 100 mK, the two are close enough to each other, that we can no longer be sure that the solid is the bottleneck to the flux near 100 mK.

The possibility that the flux is limited by the Vycor instead of the solid near 100 mK is further demonstrated by figure \ref{fig:flow-highT}b,  which shows the averages of the warming slopes, $AC = (|A|+|C|)/2$, and separately the average of the cooling slopes, $BD = (|B+|D|)/2$, for series sh03 and sh04 when increasing TC.  At higher temperatures, that is $TC \geq 200$ mK, it is seen that $AC = BD$.  But, as the temperature is lowered below 200 mK it is seen that $BD > AC$ meaning that the flux measured when cooling a reservoir is higher than when warming the reservoir for TC near 100 mK.  This can be explained by assuming that near 100 mK the Vycor is limiting the flux because at higher reservoir temperatures $\rho_s$ is reduced in the upper region of the Vycor rods (as is also seen in figure 5).

\subsection{flow for $T \leq$ 100 mK}
\label{sec:flux:lowT}
\begin{figure}
\resizebox{3.5in}{!}{
\includegraphics{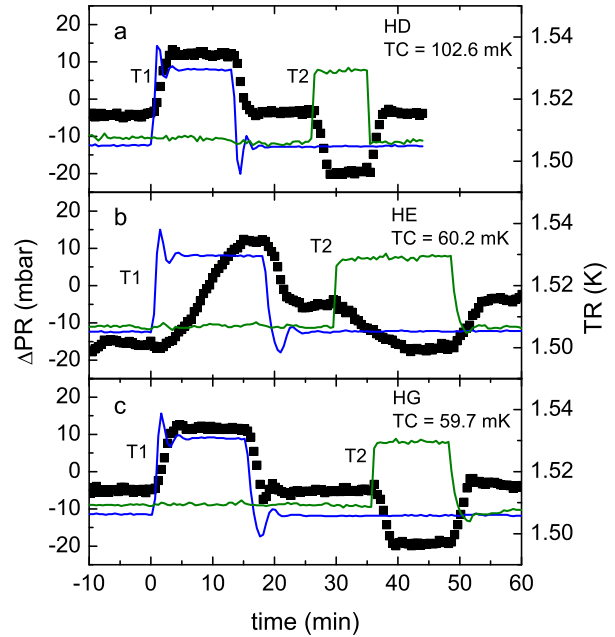}}
\caption{\label{fig:HD-HE-HG} (color online) Flux through samples (a) HD, (b) HE and (c) HG.  The clock is reset for each data set so that T1 is increased at $t = 0$.  In sample HE, which was cooled following the measurements on HD, there was much weaker coupling between the two reservoirs; the flux was much smaller.  Following a ``syringe injection" and a liquid helium transfer, sample HG showed a flux value similar to that seen for sample HD.  }
\end{figure}
Below 100 mK, some additional features are seen in the observed flux through the solid.  First, recall that series sh03 was grown at $TC = 317$ mK, then cooled to 60 mK after which the flux was measured in 100 mK increments up to 700 mK, then cooled with the flux measured, following steps in TC of 100 mK.  Near 100 mK (sample HD) the flux was consistent with earlier measurements (figure \ref{fig:HD-HE-HG}a).  For example,  for sample HD at 102.6 mK (figure \ref{fig:HD-HE-HG}a) the average of flow rate is $ABCD = 0.1175$ mbar/s.  But, when the solid was returned to 60 mK (sample HE, figure \ref{fig:HD-HE-HG}b) a smaller and variable flow was seen between the two reservoirs.  This weaker coupling is evident at the beginning of the measurement where $\Delta PR = -15$ mbar even though $T1 = T2$. Further, it is seen that the average flux is significantly smaller; for sample HE the flux is  $0.0431$ mbar/s, but the flow rate for each slope varies.  At this point a helium transfer was needed.  After filling the liquid helium bath, a ``syringe injection" \cite{Ray2010a} added atoms to the cell to raise the pressure of the solid from 25.661 bar (sample HF) to 26.053 bar. At that point a fountain measurement revealed that a higher flux, $ABCD = 0.1067$ mbar/s had returned (sample HG) (figure \ref{fig:HD-HE-HG}c).

\begin{figure}
\resizebox{3.5in}{!}{
\includegraphics{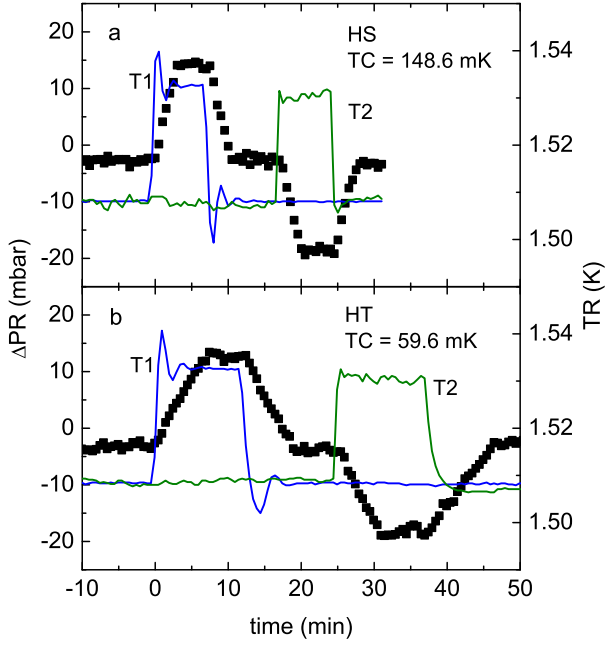}}
\caption{\label{fig:HS-HT} (color online) Samples (a) HS at $TC = 148.6$ mK and (b) HT at $TC = 59.6$ mK.  Both are part of the cooling data sets of series sh04.  After cooling to 59.6 mK, the flux dropped similar to series sh03 (figure \ref{fig:HD-HE-HG}).  Note that the time axis is shifted so that T1 is increased at $t = 0$ in each case.}
\end{figure}
After repeating the temperature cycle at the higher pressure, $C_{avg} = 26.0$ bar (series sh04) the same observations were made, i.e. when returning the sample back to $TC = 59$ mK from above 100 mK, the flux dropped.  Figure \ref{fig:HS-HT} shows the final two samples in this series, HS at $TC = 148.6$ mK, and HT at $TC = 59.6$ mK.  Just as in the previous series, there is a significant drop in the flux as the temperature is lowered to 59.6 mK, but in this case the reduced flux was more stable.

\begin{figure}
\resizebox{3.5in}{!}{
\includegraphics{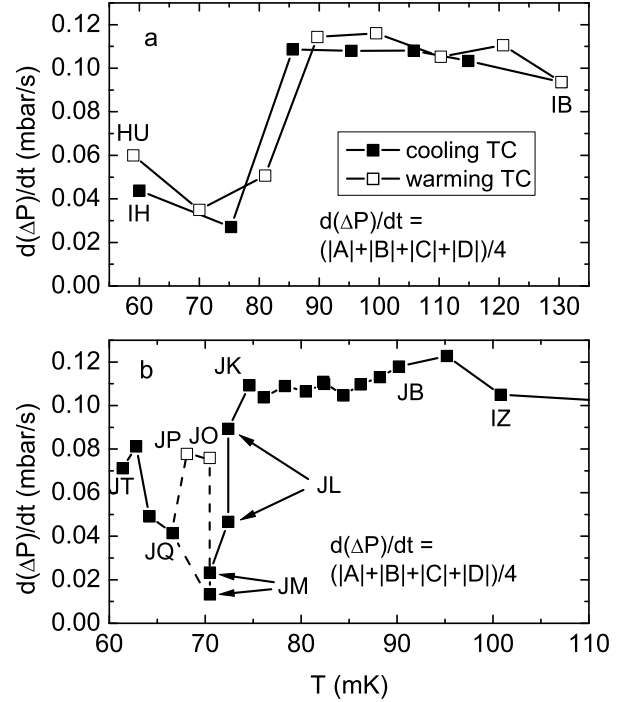}}
\caption{\label{fig:lowT} (a) series sh04b in which the flux was measured starting at $TC \approx 60$ mK, then warmed in increments of 10 mK to 130 mK, then cooled back down in 10 mK increments to 60 mK. A change in the flux is seen to occur near $TC = 80$ mK.  (b) series sh06b.  Here the flux was measured in 2 mK steps as TC was lowered from 100 mK to 60 mK, and the flux dropped in the vicinity of $TC = 75$ mK.  See text for a discussion of samples JO and JP.}
\end{figure}
Following measurement HT, a third sequence of measurements was performed with the same sample of solid helium in an effort to document this drop in the flux more closely.  This was series sh04b, and is shown in figure \ref{fig:lowT}a. Here, TC was warmed from 59 mK to 130 mK then cooled back to 60 mK with the flux measured at a stable temperature every 10 mK.  Upon warming the solid there is a clear rise in the flux, which occurs near $TC = 80$ mK, and when cooling the sample, the flux drops near the same temperature. This change in the flux between fast flow and slow flow is seen to be rather abrupt, here seen to take place within the 10 mK change in cell temperature.

After series sh04b, the solid was melted, and a new sample of solid helium was grown from the superfluid at $TC = 317$ mK, then cooled to $TC = 100.8$ mK (sample IZ).  The sample was then cooled further to $TC = 61.4$ mK with the flux measured every $\approx 2$ mK; series sh06b, figure \ref{fig:lowT}b.  Here, the flux remained high until $TC \approx 75$ mK, then dropped to near 0 at $\approx 70$ mK before rising again as the temperature was further lowered toward 60 mK.

\begin{figure}
\resizebox{3.5in}{!}{
\includegraphics{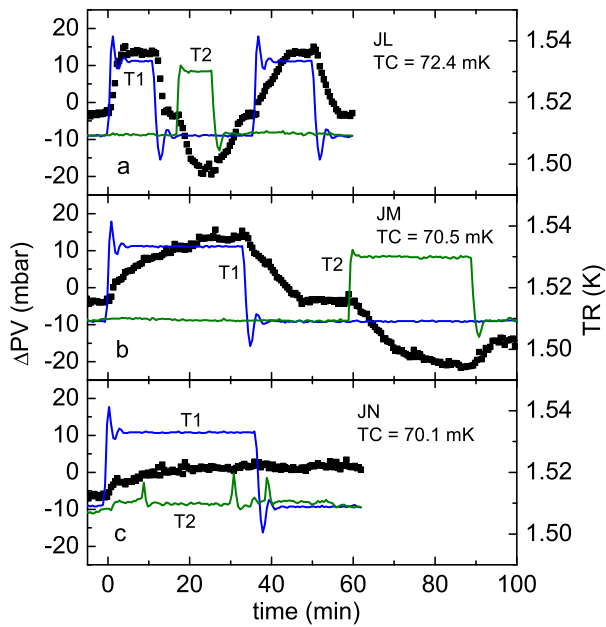}}
\caption{\label{fig:JLJMJN} (color online) Measurements (a) JL, (b) JM, and (c) JN.  Successive measurements that show strong dynamic behavior.  The time axis is shifted so that the increase in T1 occurs at $t = 0$ in each case.}
\end{figure}
 In the vicinity of the flux minimum, dynamic behavior was observed.  For example, following measurement JK at $TC = 74.6$ mK, the cell temperature was lowered to $TC = 72.5$ mK, and measurement JL was made, shown in figure \ref{fig:JLJMJN}a.  Here the first step in T1 yielded slopes $A1 = 0.100$ mbar/s and $B1 = 0.154$ mbar/s.  Then, the T2 step produced the slopes $C = 0.066$ mbar/s and $D \approx 0.034$ mbar/s, which is significantly lower than the flow rates measured in the T1 step (the D slope here is found by fitting to the steepest section).  A subsequent second step in T1 produced the slopes $A2 = 0.030$ mbar/s and $B2 = 0.054$ mbar/s, which are similar to the C and D slopes.  There was apparently some dynamic change in the solid that caused the slopes to change with time.  These two different flow rates for sample JL are shown in figure \ref{fig:lowT}b as $(|A1|+|B1|+|C|+|D|)/4$, and $(|A2|+ |B2|+|C|+|D|)/4$.

Following sample JL, TC was lowered to 70.5 mK, and measurement JM was performed (figure \ref{fig:JLJMJN}b).  The initial step in T1 produced a flow but no readily identifiable A slope, in fact, over the modest time span involved the flow data can be reasonably fit to an exponential relaxation with a time constant $\approx$ 13 min.  Here, the A flux spans a range given by  $A$ = 0.031 to 0.008 mbar/s.  We may also characterize the C-slopes by a range, $C$ = 0.022 to 0.006 mbar/s.  When T2 was subsequently returned to its original value, $|\Delta P|$ began to fall, but then stopped, and and the data collection ended (the small bit of flow seen was used to determine the D slope for fig.~\ref{fig:lowT}b).

After 1.3 hours, another fountain measurement was attempted at $TC = 70.1$ mK (sample JN, fig.~\ref{fig:JLJMJN}c).  Here, an increase in T1 produced an initial change in $P1 - P2$, but only by 5 mbar. (Using equation \ref{eq:ftn}, a 15 mbar pressure difference was expected).  Additionally, when T1 was returned to its initial value, there was no change in P1 - P2 during an observation time of 35 min.  JN is not shown in fig.~\ref{fig:lowT}b because there was time to change only T1 due to the need for a liquid helium transfer to the liquid helium bath.  Near 70 mK there clearly seems to be dynamical behavior in the system that causes the flow rate through the solid to change with time and diminish to zero.

Following data set JN, the helium bath was filled (which introduces vibration), and the apparatus was left undisturbed for $\approx 16$ hours before measurement JO, also done at $TC = 70.5$ mK, which is indicated by an open square in fig.~\ref{fig:lowT}b.  The average flux was $ABCD = 0.0755$ mbar/s, which is a marked increase from the previous few measurements.  The next measurement, JP (the other open square in figure \ref{fig:lowT}b), at $TC = 68.2$ mK showed a similar high flux, with $ABCD = 0.0773$ mbar/s.  Then, as the sample was cooled further to 66.6 mK (sample JQ), the flow rate decreased again, before rising with decreasing temperature down to $TC = 61.4$ mK.  Samples JO through JQ were all measured within $\approx 3.5$ hours of each other (1.75 hours elapsed between the beginning of JO and the beginning of JP, while 1.6 hours elapsed between the beginning of JP and of JQ).  We emphasize here that the variable flow rates seen deep in the flux valley have not been observed in other temperature regimes.

\subsection{detailed behavior of one solid sample}
\label{sec:flux:sample}
It may be useful to follow the sequential evolution of the behavior of a single solid sample from its initial growth to the completion of its study, when it was melted. Before doing so, we first describe the creation and some manipulation of the solid sample.  It was initially prepared by growth from the superfluid at 300 mK.  Following growth, a syringe injection through lines 1 and 2 was done (we denote this as sample GP) at 317 mK to elevate the pressure to about 25.8 bar.  A fountain measurement was then done (measurement GQ) at 317 mK, where the flux was found to be 0.0463 mbar/sec, a number shown in figure \ref{fig:flow-highT}.  The sample was cooled to 61 mK (measurement GR) and there followed a series of measurements (GR - HE) made while the cell temperature was increased and then decreased. Two of these measurements were at 60 mK.  Measurement GR resulted in a flux of 0.1252 mbar/sec, but measurement HE resulted in a flux of 0.0431 mbar/sec.  Measurement HE appeared anomalous. It's flux did not fit on the trend expected from the data shown in figure \ref{fig:flow-highT}.  A helium transfer was done and another syringe injection (which of course adds atoms to the sample and changes the sample pressure a bit) was done (sample HF), which resulted in a flux measurement HG (at $C_{avg}$ = 26.03 bar at 59 mK) of 0.1067 mbar/sec.

\begin{figure}
\resizebox{3.5in}{!}{
\includegraphics{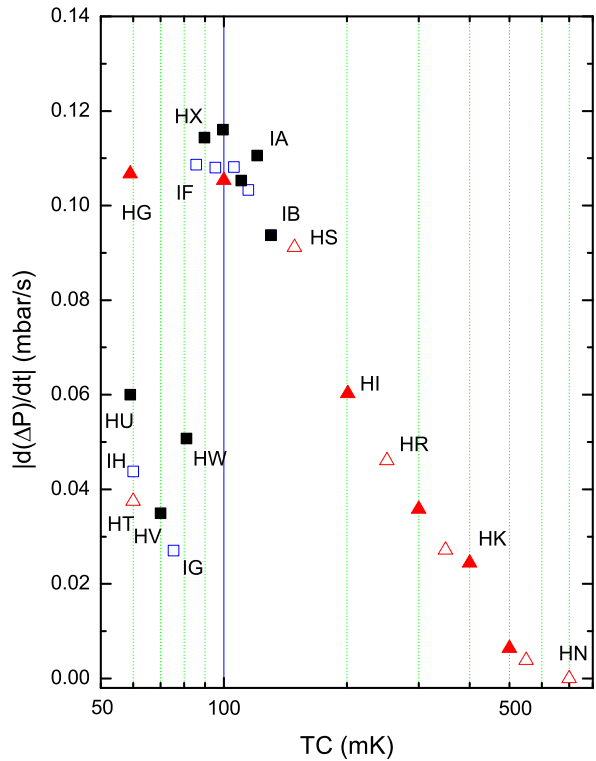}}
\caption{\label{fig:logplot} (color online) Flux behavior of a single sample of solid.  Beginning with sample HG (series sh04, sh04b), the temperature of the cell was changed to new stable values after which the flux was measured and after each change the same solid sample was given a new designation (see text and Appendix).  Here $d(\Delta P)/dt = ABCD = (|A| + |B| + |C| + |D|)/4$.  Samples HG-IH. Solid (open) symbols represent data taken while warming (cooling).}
\end{figure}

Next, consider figure \ref{fig:logplot} where we show the sequential evolution of behavior of the single solid sample measurement from HG to the end of the sequence, IH (series sh04 and sh04b), at which point a transfer was needed and the sample was melted. This plot is absent two data points for which there were apparatus problems, but the evolution is clear.  HG was warmed to 100 mK, a fountain measurement taken, and thereafter measurements were taken following 100 mK warming steps to 700 mK, (measurements HG-HN), a helium transfer was done, and the same sample was cooled to 550 mK and thereafter to 60 mK in steps of $\approx$ 100 mK (measuremens HO-HT). Measurement HT resulted in a low value for the flux.  At that point a transfer was done and the sample was measured at 59 mK (HU), where it showed a modestly higher flux, but still low. Then it was warmed in steps of 10 mK to 130 mK (measurements HV-IB), which caused the flux to pass through a minimum and a maximum and reproduce previous values of the flux in the higher temperature part of this range. The sample was then lowered in temperature to 60 mK (measurements IB-IH), passing though a flux minimum, at which point the sample was melted.  The relatively robust nature of the temperature dependence of the flux for this single solid sample is demonstrated in figure \ref{fig:logplot}.

\begin{figure}
\resizebox{3.5in}{!}{
\includegraphics{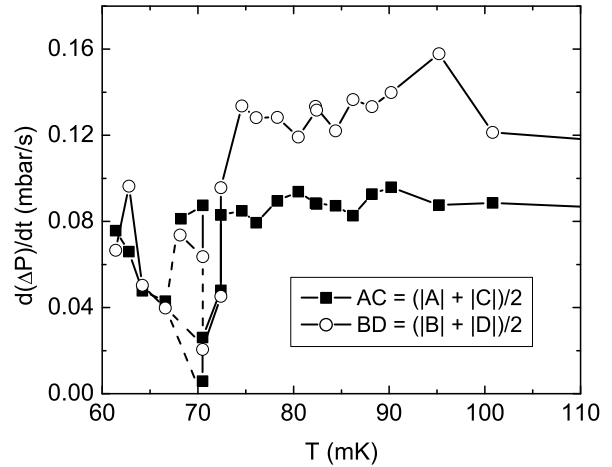}}
\caption{\label{fig:sh06b_AC-BD} The average of flow rates A and C (warming the reservoir temperature), and B and D (cooling the reservoir temperature) for series sh06b, described in section IV C.}
\end{figure}
Given the behavior of the flux below 100 mK seen in figures \ref{fig:lowT} and \ref{fig:logplot}, it is important to ensure that this minimum in the flux is not due to some property of flow through the Vycor rods.  Figure \ref{fig:sh06b_AC-BD} shows data from series sh06b described previously and shown in figure \ref{fig:lowT}b.  Here, we separately plot the average flux found after warming each reservoir, $AC = (|A| + |C|)/2$, and after cooling each reservoir, $BD = (|B| + |D|)/2$, as also shown for $T \geq 100$ mK in (figure \ref{fig:flow-highT}b).  Above the flux minimum, $TC > 73$ mK, it is seen that $BD > AC$ meaning that in {\it this} region (75 mK $<$ TC $<$ 110 mK), the Vycor is likely limiting the flow when a heater is on, just as was seen in figure \ref{fig:flow-highT}b.  However, in the vicinity of the flux minimum, it is seen that $AC \approx BD$ which indicates that the Vycor is not limiting the flow in this region, and that the flux minimum is due to some change in the flow paths in the solid chamber.

It should be noted that two solid helium samples did not show a drop in the flux near 75 mK.  The first, series sh09, was grown from the superfluid at $TC = 320$ mK and cooled directly to 60.6 mK, then the flux was measured while warming the solid in $\approx 4$ mK steps.  No change in the flux was measured with $ABCD \approx 0.10$ mbar/s up to $TC = 91$ mK, the highest temperature studied in this sequence.  Another series, sh11, involved a solid sample that was grown by the blocked capillary method and cooled to 63 mK and studied in the range 63 - 70 mK, which also did not show a drop in the flux.  The sample temperature was raised to 300 mK and a syringe injection done to increase the pressure and the sample was returned to 61.2 mK, with a flux like that seen in the 63 - 70 mK range for samples in series sh09. This indicates that the flux minimum is likely sample-dependent and not caused by a simple thermal cycle of the solid.

\section{discussion}
\label{sec:disc}
It is not clear what the specific cause of the flux through solid helium is and in particular what is causing it to decrease and then rise at the lowest temperatures accessible to us.  In samples that show the minimum in flux, it is quite robust.  For instance, the minimum was observed in series sh03, sh04 and sh04b which were all with the same solid sample that had been thermally cycled to 700 mK twice.  In these series, the samples were also left overnight at 700 mK for $\approx$ 14 hours.  Samples in series sh06, created at 317 mK and cooled to near 60 mK also showed the flux minimum.  However, the fact that two other samples did not show this means that the presence of the flux minimum is not an intrinsic property of the solid; rather it must be related to the details of each particular solid, presumably due to the presence of some form of disorder.

\subsection{nonlinearity}
It has been suggested \cite{KuklovPC,AleinikavaUP} that the observed minimum in flux through the solid could be due to non-linear effects caused by the size of the chemical potential difference applied across the solid.  If this is the case, when a sample is in the flux minimum different sized differences in the reservoir temperatures, $\delta T = T1 - T2$, are expected to produce different values for the flux.  More specifically, smaller steps in the chemical potential are predicted to produce a higher flow rate.

\begin{figure}
\resizebox{3.5in}{!}{
\includegraphics{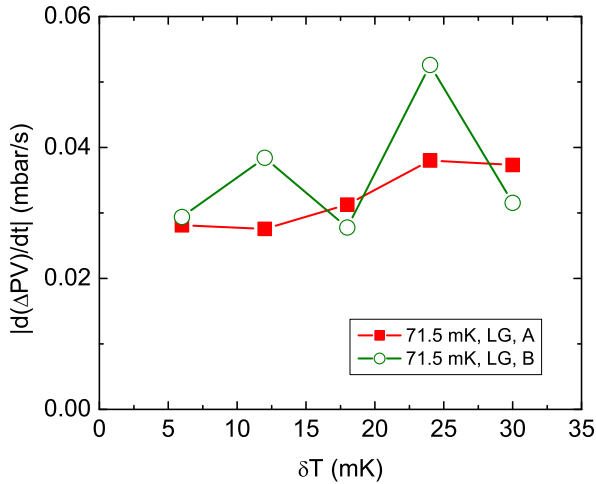}}
\caption{\label{fig:KG-LG-3}(color online) The average of flow rates A (warming T1), and B (cooling T1) for measurement LG at $TC = 71.5$ mK ($T_0 = 1.507$ K), in the region where the flux was seen to drop.}
\end{figure}
To search for the possibility of these non-linear effects, different sized temperature differences, $\delta T$, were applied between the two reservoirs.  For each $\delta T$, T1 was raised from a base temperature $T_0$ and held constant until equilibrium was established then returned to the original temperature (T2 was held constant the entire time).  In this way, the A (warming) and B (cooling) slopes could both be determined for each size step.  The results of these measurements are shown in figure \ref{fig:KG-LG-3}, which shows the flow rate as a function of the temperature step for measurements of sample LG.   Sample LG was grown from the superfluid (at T = 327 mK), then cooled directly to 61 mK, then warmed to 71.5 mK where it was seen to be in a low flux state.  Sample LG (because it was a low-flux sample) thus provides a test sample to search for non-linear effects in the flow due to the size of the chemical potential difference imposed across the solid.  Sample KG (a high flux sample at 100 mK) provides a comparison (See appendix).  For sample LG it is seen that slopes $A \approx B$, and there is no rise in the flux for smaller-sized temperature steps.  So if there are non-linear effects in the solid causing the flow rate to decrease, then they must only diminish (and the flux rise) for temperature steps smaller than 6 mK.

\subsection{possible roles of the \3he impurity}
Another possibility to consider in the context of the drop in the mass flux (e.g. at $\sim$ 75 mK; figure \ref{fig:sh06b_AC-BD} ) might be that it is due to the $^3$He-induced pinning of dislocations.  Pinning of dislocations is the mechanism that is thought to be responsible for the observed increase in the shear modulus with decreasing temperature first reported by Day and Beamish \cite{Day2007}, and the softening with increasing temperature seen by Rojas {\it et al.} \cite{Rojas2010a}.  As the temperature is reduced, the dislocations become pinned either to each other, or to defects such as \3he impurities, which makes the solid stiffer.  If the dislocations provide the superfluid paths through solid helium, it is possible that the pinning can either enhance the effective conductivity of the pathways (pinning to each other) or reduce the flow of \4he atoms through some of the superfluid cores (pinning by \3he condensation on dislocation cores).  Pinning by \3he might be a cause for the drop in flux because the presence of \3he on a dislocation might have an effect that would decrease the flux of \4he along the dislocation core. And, it is possible that further stable crosslinking of the dislocations as the temperature is further reduced might provide bypass pathways around the blocking $^3$He allowing a flux increase across the sample, while at the same time stiffening the sample.

We believe that it is unlikely that the drop in flux is in some (unknown) way related to bulk \3he - \4he phase separation.  With the commercially available helium used in these experiments the concentration of \3he is $x_3 \sim 300$ ppb.  For this concentration the phase separation temperature is expected to be $T_{PS} = 51$ mK \cite{Edwards1989}.  It is not clear, however, if this is the actual concentration of \3he in the solid.  The cell is initially filled with helium from the third fill line, which bypasses the Vycor and preserves the \3he concentration.  The solid is most often grown above the melting curve by addition of \4he through the Vycor rods, which presumably filters out the \3he. This would tend to lower the $x_3$ in the solid, and lower the phase separation temperature. It is also possible that the \3he in the cell diffuses to liquid regions, during solid growth \cite{Pantalei2010}, which may be present in the cell (perhaps including the Vycor, which is filled with liquid).  This would further lower $x_3$ in the solid and  the phase separation temperature of 51 mK should be considered as an upper limit; $T_{PS} \lesssim 51 $ mK.  This makes it unlikely that phase separation is in any way associated with the observed low temperature drop in flux near 75 mK.

\subsection{injection {\it vs} fountain measurements}
Connections can be made between these flow experiments, and our previous flow experiments where a mass flow was induced by the injection method \cite{Ray2008a,Ray2009b}.  In those experiments, flow was seen to cease for temperatures greater than $\approx$ 600 mK, which is similar to the observations noted here where no flow was seen for $TC > 650$ mK.  Also, in both experiments, the mass flux is seen to be independent of the applied chemical potential gradient, which is consistent with superflow at a critical velocity.

There are also some differences between these results using the fountain effect to induce flow, and the previous injection experiments.  Most notable is the absence of hysteresis when the temperature is thermally cycled using the fountain method.  In the injection experiments, when a sample that showed mass flow was warmed to $T \gtrsim 600$ mK where it did not flow, it usually did not flow when cooled back to the original temperature (unless, once at that temperature the cell pressure was reduced slightly; then flow could be initiated).  This behavior was not seen when using the fountain method.  We note, however, that the procedure for the temperature cycles was somewhat different between the injection and fountain methods.  In the injection methods, the sample was typically warmed in one step from low temperature (usually near 400 mK) to $T > 600$ mK, then returned with a temperature step of a similar size.  In the fountain method, the sample was warmed in several smaller steps, but only to the temperature where flow was no longer observed, then cooled back down in steps.

Aside from slight differences in measurement protocol, the reason for the difference in behavior between the two methods used to induce flow may have to do with the fragility of helium crystals.  It has been seen that in some temperature regimes solid helium crystals can be remarkably fragile, with non-linear effects occurring in strains\cite{Day2007,Pantalei2010} as low $\sim 4 \times 10^{-8}$ , which corresponds to a stress\cite{Day2010} of $\sim 8 \times 10^{-6}$ bar.  The smallest pressure steps applied in the injection experiments was $\sim 0.1$ bar.  This makes it very likely that each injection changes the solid in an irreversible way so that the next measurement has a different set of starting parameters.  Using the fountain method, the pressure differences applied between R1 and R2 are typically $\sim 7 \times 10^{-3}$ bar.  This is still a substantial pressure difference, but in this method the total mass in the system stays the same. By returning the Vycor temperatures back to their original values, each measurement can start with the same base pressure.
We note here that whether by injection or by fountain techniques, when a change in cell pressure is seen in C1, a similar change is seen in C2. This behavior is consistent with the climb of edge dislocations.

\subsection{flow scenarios}
Although as we have mentioned, it is not possible to make a definitive statement about what mechanism is at work to carry the flux we observe, it is useful to consider some of the possibilities that have been proposed.  To do so in a quantitative way, we take a typical flux value of 0.06 mbar/sec, which corresponds to a typical mass flux (through the solid from one reservoir to the other) of 2.9 $\times 10^{-8}$ g/sec.

In the case of bulk flow we can write the mass flux as $dm/dt = \xi v\rho A$, where $\xi$ represents the supersolid fraction, v the velocity, $\rho$ the density, and A the cross sectional area that carries the flow.  With a channel diameter of 0.64 cm, and a typical mass flux of $dm/dt$ = 2.9 $\times 10^{-8}$ g/sec we find that $\xi v = 5.1 \times 10^{-7}$ cm/sec.  So, if we take $\xi$ = 0.01, then $v$ = 5.1 $\times 10^{-5}$ cm/sec or 0.54 $\mu$/sec. On the other hand, if we take a velocity characteristic of that implied by some of the torsional oscillator experiments, v = 10 $\mu$/sec, then we find, $\xi$ = 5.1 $\times 10^{-5}$, which is smaller than the typical values deduced from torsional oscillator experiments, $\xi$ = 0.01 $\%$ to a few percent.

Edge dislocations have been suggested as a conduit that would allow superflux and also provide a mechanism for the increase in the density of the solid that is seen in our experiments.  If we think in terms of an edge dislocation, the edge of the dislocation will carry the flux. To be quantitative, we assume that an edge dislocation has an effective cross sectional area of $A$ = 1 nm$^2$.  In that case, with $dm/dt = \xi vN\rho A$ = 2.9 $\times 10^{-8}$ g/sec, where $N$ is the number of edge dislocations that contribute to the flow, we find that $\xi vN = 1.7 \times 10^{7}$ cm/sec. If we take $\xi$ = 1, and adopt a velocity of v $\leq$ 350 m/sec (typical of a first sound velocity) as the maximum possible flow velocity, then N $\geq$ 480, which suggests a dislocation density of $\geq$ 1600 cm$^{-2}$. For a smaller value of $v$ there is a proportionate increase in N.

It was suggested previously \cite{Balibar2008a} that our flow results could be caused by liquid channels which have been shown to exist in solid helium samples on the melting curve \cite{Sasaki2007}.  We have argued for various reasons \cite{Ray2008e,Ray2009b} that liquid channels are not likely the cause of our observed mass flux.  The expected cross sectional area of a liquid channel\cite{Sasaki2008} at 26 bar is $\approx 40$ nm$^2$.  Interpolating data\cite{Lie-zhao1986,Shirahama2007a} gives a superfluid transition for this size channel of $\approx$ 1.55K.  This is inconsistent with our observations, which show that flow ceases at around 700 mK. None the less, we can apply a similar quantitative discussion and write that in the case of $N_L$ liquid channels each of diameter $A \approx 40$ nm$^2$, the mass flux is given by $dm/dt = \xi vN_L\rho A$. So, with $dm/dt$ = 2.9 $\times 10^{-8}$ g/sec we have $\xi vN_L = 4.3 \times 10^{5}$ cm/sec. If we take $\xi $ = 0.1, then $vN = 4.3 \times 10^{6}$ cm/sec. If we make the assumption that what limits the flow in such a channel is a critical velocity given by $v_c = K/2R$, where $K$ is the circulation and $R$ = 7 nm, then we find $v_c = 1.4 \times 10^{3}$. This then results in $N_L$ = 3070 liquid channels.  Although there are a number of assumptions, this strikes us as a rather large number.

Perhaps the most convincing argument that liquid channels are not relevant is the observation that if liquid channels were present, it is hard to understand how the flux for superfluid in such channels would decrease at lower temperatures ($\approx$ 75 mK) and then increase again at even lower temperatures. So, we continue to believe that the evidence does not favor liquid channels as the conduction mechanism for the mass flux we measure.

\subsection{comparisons to other work}
The relation between these flow experiments, the torsional oscillator experiments and the shear modulus experiments remains unclear, although the dramatic changes in flux behavior that we measure take place near the temperature at which the torsional oscillator period and the shear modulus change rapidly with temperature.  But, our experiments have shown the onset of flow to begin near $T \approx 600$ mK; the resonant period drop in the torsional oscillators (and rise in shear modulus) does not begin to change until 100-200 mK depending on the parameters of the sample, with the most dramatic changes seen near 70 mK for solid samples made with commercial \4he.

One possibility is that the effects in the torsional oscillators at low temperatures are there at higher temperatures, but not visible at the frequencies at which the oscillators operate ($\sim$ 500 - 1000 Hz); our present measurements are at essentially zero frequency. One might imagine that if dislocations are responsible, configurational changes might influence the tortuosity\cite{Johnson1982, Wong1993}. Increases in tortuosity might make superfluidity be less visible at higher temperatures in torsional oscillator experiments, but this seems unlikely for a macroscopic dislocation density. If Reppy\cite{Reppy2010} is correct, the torsional oscillator period shift seen at low temperatures is not related to supersolidity.  It may be that the presence of flow for T $>$ 80 mK is due to different sample behavior than that which is causing the flow at lower temperatures.  It is possible that the torsional oscillators observe the stiffening transition, while we see a true mass flux, a flux that is modified by the presence of the stiffening transition.

\subsection{comment}
The presence of dynamic behavior in the vicinity of the flux minimum causes us to offer a speculation.  Perhaps it is possible that for T $\gtrsim$ 100 mK many mobile dislocation lines can carry a net mass flux.  For T $\lesssim$ 70 mK a relatively stable network of cross linked dislocations may also carry a net flux. But, in the vicinity of 75 mK there may be fluctuations in cross-linking, which interrupt the effective flux in this crossover region, with perhaps a role for \3he as noted earlier, and this results in the minimum. Perhaps this crossover region is what is seen in the specific heat measurements. In this scenario, the torsional oscillators and shear modulus experiments see a stiffening transition and the effects of flow are only seen in the experiments we have carried out, and perhaps the DC rotation experiments\cite{Choi2010}.

\section{summary}
\label{sec:sum}
In section \ref{sec:fountain} we reported the observation of a thermo-mechanical (or fountain) pressure difference between two liquid reservoirs separated by hcp solid helium pierced by two spacially separated Vycor rods.  This observation can be interpreted to mean that there is indeed a pathway (or pathways) for mass to flow through solid helium.  Further, the constant flux, independent of pressure difference, through solid helium shown in section \ref{sec:flux} is indicative of a superflow at critical velocity, and so provides evidence that the paths are likely superfluid.  Thus, taken as a whole, these two observations support the conclusion that superfluid-like behavior is present in a cell filled with solid \4he.  This superfluidity in solid helium is seen to appear at $T \approx 700$ mK, and the flux increases as the temperature is decreased until $\approx$ 100 mK.  As the temperature is further lowered below 100 mK, the flux likely continues to rise (though we were limited in the maximum flux we could measure).  This increase in the mass flux may signify an increasing superfluid fraction, $\rho_s$, in the conducting pathways or an increase in the effective number of conducting pathways.

It is still not clear what these pathways that allow for a superfluid connection through the solid are.  Though much theoretical attention has been directed lately to the superfluidity of dislocations \cite{Boninsegni2007,Pollet2008}, there are other possibilities, such as flow through grain boundaries \cite{Pollet2007} and flow through a glassy phase of solid helium \cite{Boninsegni2006}.  Unfortunately, our apparatus cannot conclusively distinguish among these scenarios. But, the possibility of dislocations with dislocation crosslinking and perhaps \3he condensation, which seem consistent with the features we observe, appears to us to be the strongest possibility.

Near $TC = 75$ to $80$ mK a sharp decrease in the flux is seen, with a subsequent rise the temperature is further lowered.  It is not clear what causes this feature, but we have offered a possible scenario.  It appears as though this drop is {\it not} an intrinsic property of the solid since some samples did not show a drop in the flux.  We think it most likely that the flux minimum has something to do with the connectivity, \3he condensation and cross-linking of dislocations in the solid. We have not completely ruled out the presence of non-linearity that might be caused by the size of the applied temperature differences \cite{KuklovPC}.  Finally,  the flow rate is seen to rise again as the temperature is further lowered, which may be a sign of a switch in the mechanism that is causing the flow.  Further study of this phenomenon to lower temperatures, higher pressures and different \3he concentrations is necessary to fully understand what is causing it.

\section{acknowledgments}
We have benefitted from numerous interactions with B. Svistunov over the course of evolution of our experiments. He and N. Prokofev have been stimulating colleagues throughout our work.  We have also had helpful conversations with many others including S. Balibar, J. Beamish, M.C.W. Chan, H. Kojima, A. Kuklov, W. Mullin and J. Reppy. This work was supported by the National Science Foundation primarily through National Science foundation grant DMR 08-55954, with some support from DMR 07-57701 and with initial support from DMR 06-50092. We have also benefitted from incidental support from Research Trust Funds provided by the University of Massachusetts Amherst and access to facilities provided by the MRSEC (NSF DMR 08-20506) at the University.

\bibliography{ref}

\section{appendix 1}
\begin{figure} [b]
\resizebox{3.5in}{!}{
\includegraphics{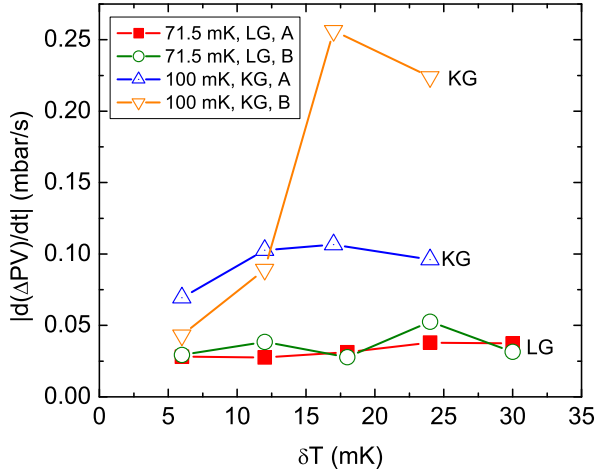}}
\caption{\label{fig:KG-LG}(color online) The average of flow rates A (warming T1), and B (cooling T1) for measurement LG at $TC = 71.5$ mK ($T_0 = 1.507$ K), in the region where the flux was low, and for measurement KG at $TC = 100$ mK ($T_0 = 1.509$ K).}
\end{figure}
In addition to the search for on-linear effects in a low flux sample (sample LG), we did a comparison study of a high flux sample, KG. Sample KG was a solid helium sample grown from the superfluid at T = 352 mK, then cooled to 100 mK where the different-sized temperature steps were applied. In sample KG for $\delta T \gtrsim 18$ mK it is seen that slope $B >$ slope $A$, which is expected since the B slopes are measured at lower reservoir temperatures (figure \ref{fig:KG-LG}). The apparent drop in the B slopes for sample KG for $\delta T \lesssim 12$ mK is likely due to the time constant associated with the reservoir's temperature controller, due to its design.For smaller applied $\delta T$ the temperature controller changes the heater input at a slower rate, and since at 100 mK sample KG was in the high flux state, the flux is actually being limited by the rate at which the power delivered by the heater is increased in this small $\delta T$ regime.
\begin{figure} [t]
\resizebox{3.5in}{!}{
\includegraphics{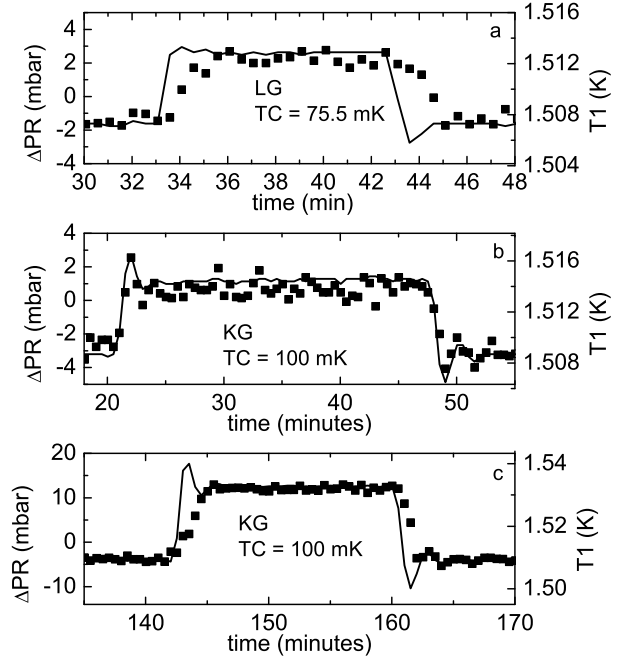}}
\caption{\label{fig:KG-LG-time} (a) $\Delta PR$ for the 6 mK step in sample LG (b) $\Delta PR$ for the 6 mK step in sample KG, and (c) $\Delta PR$ for the 24 mK step in sample KG.}
\end{figure}

Figure \ref{fig:KG-LG-time} shows an expanded view of data measured with $\delta T = 6$ mK for samples LG (a) and KG (b), and $\delta T = 24$ mK for sample KG (c). (We note here that for all of the solid \4he data we have presented prior to this point, through figure \ref{fig:sh06b_AC-BD}, $\delta T = 24$ mK.)  Here it is seen that for the 6 mK step in KG (figure \ref{fig:KG-LG-time}(b)) the time for the temperature to equilibrate, $\tau_T$, is $\tau_T \approx \tau_P$, where $\tau_P$ is the time for the pressure equilibrate. In other words, the rate of change of the reservoir temperature is likely limiting the observed flow in the case of the 6 mK steps in reservoir temperature for sample KG.  On the other hand, it is seen that for the 6 mK step in LG (figure \ref{fig:KG-LG-time}(a)), $\tau_T < \tau_P$, as is also the case for larger steps in sample KG, as shown in figure \ref{fig:KG-LG-time}(c).  This means that the discussion in section VA is correct, and in the low flux state no rise in the flux is seen at temperature steps as low as 6 mK (chemical potential difference = 4.8 $\times$ 10$^{-10}$ J), i.e. non-linear effects are not observed.  Further, if the minimum in the flux is due to non-linearities caused by the size of the temperature step (meaning that $\delta T < 6$ mK is needed), then this problem of equilibration times will make the effect very difficult to see in this apparatus as currently configured, since the situation will not improve as $\delta T$ becomes smaller.

\section{appendix 2}
In this appendix we provide tables of flux values and base sample temperatures and pressures for many of the samples that are described in this report. In all of these tables, the temperature (TC) is in mK units, the pressures (C1, C2) are in bar units, and the flux values (AC, BD and ABCD) are in mbar/sec units. The units are not shown on the individual tables to improve the format.  For the flux values, the designation AC means that the absolute values of the slopes of segments A and C (measured following an increase of T1 (A) or T2 (C)) have been averaged. Similarly, the designation BD means that the absolute values of the slopes of segments B and D (measured following a decrease of T1 (B) or T2 (D)) have been averaged.  ABCD designates an average of the absolute values of all four slopes.  If a sample is missing from the alphabetical list, this signifies that (1) there was an instability of some sort in the data collection, which made the measurement unreliable or (2) the absence of a meaningful slope.

\begin{table} [b]
\caption{series sh03}
\begin{ruledtabular}
\begin{tabular}{ccccccc}
sample & TC & C1 & C2 & AC & BD & ABCD  \\
\hline
GR &  61 & 25.723 & 25.829 & 0.0996 & 0.1508 & 0.1252 \\
GS & 100 & 25.724 & 25.815 & 0.0958 & 0.1502 & 0.1230 \\
GT & 199 & 25.706 & 25.800 & 0.1036 & 0.1142 & 0.1089 \\
GU & 402 & 25.717 & 25.840 & 0.0477 & 0.0469 & 0.0473 \\
GV & 498 & 25.720 & 25.865 & 0.0289 & 0.0267 & 0.0278 \\
GW & 604 & 25.746 & 25.927 & 0.0109 & 0.0052 & 0.0080 \\
GX & 698 & 25.720 & 25.930 & 0.0000 & 0.0000 & 0.0000 \\
GY & 612 & 25.718 & 25.909 & 0.0065 & 0.0098 & 0.0082 \\
GZ & 488 & 25.714 & 25.894 & 0.0222 & 0.0278 & 0.0250 \\
HA & 415 & 25.715 & 25.897 & 0.0375 & 0.0302 & 0.0338 \\
HB & 299 & 25.761 & 25.948 & 0.0533 & 0.0616 & 0.0574 \\
HC & 205 & 25.760 & 25.944 & 0.0850 & 0.0865 & 0.0858 \\
HD & 103 & 25.759 & 25.935 & 0.1101 & 0.1249 & 0.1175 \\
HE &  60 & 25.760 & 25.937 & 0.0207 & 0.0655 & 0.0431 \\

\end{tabular}
\end{ruledtabular}
\end{table}

\begin{table}
\caption{series sh04}
\begin{ruledtabular}
\begin{tabular}{ccccccc}
sample & TC & C1 & C2 & AC & BD & ABCD  \\
\hline
HG &  59 & 25.914 & 26.152 & 0.0980 & 0.1155 & 0.1067 \\
HH & 100 & 25.913 & 26.154 & 0.0880 & 0.1228 & 0.1054 \\
HI & 201 & 25.915 & 26.164 & 0.0626 & 0.0579 & 0.0603 \\
HJ & 300 & 25.917 & 26.174 & 0.0403 & 0.0313 & 0.0358 \\
HK & 400 & 25.931 & 26.202 & 0.0270 & 0.0219 & 0.0244 \\
HL & 500 & 25.885 & 26.126 & 0.0087 & 0.0041 & 0.0064 \\
HN & 700 & 25.892 & 26.120 & 0.0000 & 0.0000 & 0.0000 \\
HO & 550 & 25.960 & 26.224 & 0.0048 & 0.0029 & 0.0038 \\
HQ & 349 & 25.993 & 26.242 & 0.0279 & 0.0266 & 0.0272 \\
HR & 251 & 25.994 & 26.239 & 0.0466 & 0.0456 & 0.0461 \\
HS & 149 & 25.995 & 26.233 & 0.0922 & 0.0901 & 0.0912 \\
HT &  60 & 25.994 & 26.223 & 0.0405 & 0.0345 & 0.0375 \\
\end{tabular}
\end{ruledtabular}
\end{table}

\begin{table}
\caption{series sh04b}
\begin{ruledtabular}
\begin{tabular}{ccccccc}
sample & TC & C1 & C2 & AC & BD & ABCD  \\
\hline
HU & 59.0 & 25.939 & 26.150 & 0.0740 & 0.0461 & 0.0600 \\
HV & 70.9 & 25.942 & 26.154 & 0.0346 & 0.0353 & 0.0350 \\
HW & 81.9 & 25.944 & 26.157 & 0.0388 & 0.0627 & 0.0507 \\
HX & 89.7 & 25.947 & 26.163 & 0.0876 & 0.1413 & 0.1144 \\
HY & 99.5 & 25.949 & 26.166 & 0.1076 & 0.1246 & 0.1161 \\
HZ & 110.3& 25.952 & 26.172 & 0.0952 & 0.1154 & 0.1053 \\
IA & 120.7& 25.953 & 26.174 & 0.1093 & 0.1119 & 0.1106 \\
IB & 130.4& 25.953 & 26.174 & 0.0898 & 0.0976 & 0.0937 \\
IC & 114.9& 25.953 & 26.174 & 0.0918 & 0.1148 & 0.1033 \\
ID & 105.8& 25.976 & 26.203 & 0.0875 & 0.1288 & 0.1081 \\
IE & 95.4 & 25.977 & 26.205 & 0.0945 & 0.1216 & 0.1081 \\
IF & 85.6 & 25.978 & 26.206 & 0.0884 & 0.1290 & 0.1087 \\
IG & 75.3 & 25.985 & 26.213 & 0.0306 & 0.0235 & 0.0271 \\
IH & 60.0 & 25.993 & 26.224 & 0.0469 & 0.0407 & 0.0438 \\
\end{tabular}
\end{ruledtabular}
\end{table}

\begin{table}
\caption{sh06b}
\begin{ruledtabular}
\begin{tabular}{ccccccc}
sample & TC & C1 & C2 & AC & BD & ABCD  \\
\hline
IY & 326 & 26.021 & 26.019 & 0.0463 & 0.0453 & 0.0458 \\
IZ & 100.8 & 26.002 & 25.979 & 0.0887 & 0.1214 & 0.1050 \\
JA & 95.2 & 25.998 & 25.974 & 0.0876 & 0.1579 & 0.1228 \\
JB & 90.2 & 26.000 & 25.976 & 0.0959 & 0.1399 & 0.1179 \\
JC & 88.2 & 26.004 & 25.981 & 0.0928 & 0.1333 & 0.11302 \\
JD & 86.2 & 26.009 & 25.986 & 0.0827 & 0.1367 & 0.1097 \\
JE & 84.4 & 26.012 & 25.991 & 0.0873 & 0.1221 & 0.1047 \\
JF & 82.3 & 26.016 & 25.996 & 0.0885 & 0.1334 & 0.1110 \\
JG & 82.4 & 26.054 & 26.030 & 0.0881 & 0.1317 & 0.1099 \\
JH & 80.5 & 26.056 & 26.040 & 0.0939 & 0.1192 & 0.1065 \\
JI & 78.3 & 26.057 & 26.040 & 0.0896 & 0.1283 & 0.1089 \\
JJ & 76.1 & 26.056 & 26.039 & 0.0795 & 0.1282 & 0.1038 \\
JK & 74.6 & 26.057 & 26.047 & 0.0849 & 0.1337 & 0.1093 \\
JL & 72.4 & 26.058 & 26.050 & 0.0480 & 0.0453 & 0.0466 \\
JM & 70.5 & 26.059 & 26.051 & 0.0135 & 0.0206 & 0.0171 \\
JO & 70.5 & 26.040 & 26.108 & 0.0871 & 0.0638 & 0.0755 \\
JP & 68.2 & 26.039 & 26.108 & 0.0811 & 0.0735 & 0.0773 \\
JQ & 66.6 & 26.041 & 26.112 & 0.0430 & 0.0399 & 0.0414 \\
JR & 64.2 & 26.042 & 26.113 & 0.0478 & 0.0504 & 0.0491 \\
JS & 62.8 & 26.220 & 26.466 & 0.0661 & 0.0964 & 0.0812 \\
JT & 61.4 & 26.200 & 26.437 & 0.0757 & 0.0666 & 0.0712 \\
\end{tabular}
\end{ruledtabular}
\end{table}

\begin{table}
\caption{series sh09}
\begin{ruledtabular}
\begin{tabular}{ccccccc}
sample & TC & C1 & C2 & AC & BD & ABCD  \\
\hline
KO & 60.6 & 25.986 & 25.936 & 0.0807 & 0.1168 & 0.0987 \\
KP & 62.2 & 26.158 & 26.078 & 0.0857 & 0.1179 & 0.1018 \\
KQ & 66.0 & 26.178 & 26.091 & 0.0837 & 0.1142 & 0.0990 \\
KR & 69.6 & 26.195 & 26.104 & 0.0846 & 0.1146 & 0.0996 \\
KS & 75.3 & 26.156 & 26.102 & 0.0842 & 0.1115 & 0.0978 \\
KU & 80.5 & 26.196 & 26.133 & 0.0745 & 0.1440 & 0.1092 \\
KV & 85.5 & 26.214 & 26.146 & 0.0872 & 0.1648 & 0.1260 \\
KW & 91.0 & 26.133 & 26.084 & 0.1088 & 0.2026 & 0.1557 \\
\end{tabular}
\end{ruledtabular}
\end{table}

\begin{table}
\caption{series sh11}
\begin{ruledtabular}
\begin{tabular}{ccccccc}
sample & TC & C1 & C2 & AC & BD & ABCD  \\
\hline
KZ & 61.2 & 25.825 & 25.814 & 0.0927 & 0.1464 & 0.1195 \\
LA & 64.7 & 25.769 & 25.783 & 0.0988 & 0.1579 & 0.1283 \\
LB & 69.0 & 25.759 & 25.775 & 0.1106 & 0.2077 & 0.1591 \\
LD & 61.2 & 25.868 & 25.906 & 0.1067 & 0.1511 & 0.1289 \\
LE & 64.5 & 25.876 & 25.909 & 0.0911 & 0.1755 & 0.1333 \\
\end{tabular}
\end{ruledtabular}
\end{table}

\begin{table}
\caption{series sh12}
\begin{ruledtabular}
\begin{tabular}{ccccccc}
sample & TC & C1 & C2 & AC & BD & ABCD  \\
\hline
LF \footnotemark[1] & 26.095 & 26.044 & 61.2 & 0.0471 & 0.0498 & 0.0485 \\
LG \footnotemark[1] & 26.103 & 26.055 & 75.5 & 0.0380 & 0.0561 & 0.0470 \\
LH & 81.9 & 26.139 & 26.088 & 0.0484 & 0.0415 & 0.0449 \\
LI & 84.7 & 26.141 & 26.109 & 0.0934 & 0.1662 & 0.1298 \\
LJ & 91.6 & 26.309 & 26.261 & 0.0806 & 0.1805 & 0.1305 \\
LK \footnotemark[1] & 26.389 & 26.332 & 94.7 & 0.1289 & 0.1588 & 0.1439 \\
\end{tabular}
\end{ruledtabular}
\footnotetext[1]{Only A and B slopes were used.}
\end{table}

\end{document}